# Absent thermal equilibration on fractional quantum Hall edges over macroscopic scale

Ron Aharon Melcer[1], Bivas Dutta[1], Christian Spånslätt[2,3,4], Jinhong Park[5], Alexander D. Mirlin[3,4,6,7], and Vladimir Umansky[1].

1. Braun Center for Submicron Research, Department of Condensed Matter Physics, Weizmann Institute of Science, Rehovot 761001, Israel
2. Department of Microtechnology and Nanoscience, Chalmers University of Technology, S-412 96 Göteborg, Sweden
3. Institute for Quantum Materials and Technologies, Karlsruhe Institute of Technology, 76021 Karlsruhe, Germany
4. Institut für Theorie der Kondensierten Materie, Karlsruhe Institute of Technology, 76128 Karlsruhe, Germany
5. Institute for Theoretical Physics, University of Cologne, Zülpicher Str. 77, 50937 Köln, Germany
6. Petersburg Nuclear Physics Institute, 188300 St. Petersburg, Russia
7. L. D. Landau Institute for Theoretical Physics RAS, 119334 Moscow, Russia

**Two-dimensional topological insulators, and in particular quantum Hall states, are characterized by an insulating bulk and a conducting edge. Fractional states may host both *downstream* (dictated by the magnetic field) and *upstream* propagating edge modes, which leads to complex transport behavior. Here, we combine two measurement techniques, local noise thermometry and thermal conductance, to study thermal properties of states with counter-propagating edge modes. We find that, while charge equilibration between counter-propagating edge modes is very fast, the equilibration of heat is extremely inefficient, leading to an almost ballistic heat transport over macroscopic distances. Moreover, we observe an emergent quantization of the heat conductance associated with a strong interaction fixed point of the edge modes. This new understanding of the thermal equilibration on edges with counter-propagating modes is a natural route towards extracting the topological order of the exotic 5/2 state.**

The quantum Hall effect (QHE) is perhaps the most studied two-dimensional topological phenomenon. Whereas excitations in the sample bulk are localized, gapless excitations flow in one-dimensional chiral modes along the edge[1]. The integer quantum Hall effect—emerging from quantization of electron cyclotron orbits and an integer occupation of Landau levels—is well understood within a single electron picture. By contrast, the richer fractional quantum Hall effect arises from strong electron-electron interactions. Among fractional states, particularly peculiar are the hole-conjugate states (at fillings $\nu = \frac{p}{2p-1}$, with integer $p > 1$, i.e., 2/3, 3/5, 4/7,…) since they host upstream propagating edge modes. Such modes have been observed for various filling factors and devices[2–5].

Topological properties of the bulk are reflected in the edge structure, allowing edge-transport coefficients to be quantized. In particular, with full charge equilibration (e.g., by impurity scattering) between counter-propagating modes, the two-terminal electrical conductance is given by $G_{2T} = \frac{e^2}{h}\nu$, where $\nu$ is the filling factor. Likewise, for thermally equilibrated edges, the two-terminal heat conductance is quantized[6,7] as $G_{2T}^Q = \kappa_{2T}T = |\nu_Q|\kappa_0 T$, where $\kappa_0 = \pi^2 \frac{k_B^2}{3h}$, with $k_B$ the Boltzmann constant, $T$ the temperature, and $\nu_Q$ is an integer (half-integer for non-Abelian states[8]). Importantly, the value of $\nu_Q$ is an inherent property of the bulk topological order. Specifically, for Abelian states, $\nu_Q$ is given by the *net* number of edge modes, $\nu_Q = n_d - n_u$, with $n_d$ ($n_u$) being the number of downstream (upstream) modes. In hole-conjugate states, $\nu_Q$ can be zero or negative, implying transport of heat in the direction opposite to the charge flow. In recent years, thermal conductance measurements were successfully performed in GaAs and in graphene[9–12], manifesting the quantization of thermal conductance for both integer and thermally equilibrated fractional states. Nonetheless, a detailed understanding of thermal equilibration on the edge is crucial for interpreting[13–16] the recent observation[11] of $\kappa_{2T} \approx 2.5\kappa_0$ at filling $\nu = \frac{5}{2}$. As both the topological order and the extent of thermal equilibration are not known, two contradicting explanations were proposed: (i) full thermal equilibration[11], which implies $\nu_Q = 2.5$, indicating a topological order known as the *PH-Pfaffian,* which is further supported by a recent experiment[17], (ii) partial thermal equilibration[15] and $\nu_Q = 1.5$, indicating the *anti-Pfaffian* order, which is supported by numerical simulations[18,19]. Our present work paves the way towards the solution, by combining local thermometry with thermal conductance measurements to study thermal equilibration in Abelian fractional states, for which the topological order is known.

The interplay between topology and equilibration also determines the temperature profile along the edge. Consider an edge with two contacts attached at $x = 0$ and $x = L$, and upstream modes sourced at an elevated temperature $T_u(0) = T_m$, while downstream modes emerge from a cold drain contact (for simplicity assumed to be at zero temperature: $T_d(L) = 0$). With efficient thermal equilibration (thermal equilibration length $l_{\text{eq}} \ll L$), the temperature of the upstream modes after propagation, $T_u(L)$, is expected to be qualitatively different for three topologically distinct classes[20] determined by the sign of $\nu_Q$: (i) for $\nu_Q > 0$ (e.g., in $\nu = 5/3$), upstream modes lose energy propagating upstream and arrive cold at the drain up to exponential corrections: $T_u(L) \sim T_m e^{-\frac{L}{l_{\text{eq}}}}$, (ii) for $\nu_Q < 0$ ($\nu = 3/5$), the injected heat propagates ballistically to the drain up to exponential corrections: $T_u(L) \sim T_m \left( const + e^{-\frac{L}{l_{\text{eq}}}} \right)$, (iii) for $\nu_Q = 0$ ($\nu = 2/3$), the thermal transport is diffusive rather than ballistic, resulting in $T_u(L) \sim T_m \left( \frac{l_{\text{eq}}}{L} \right)^{\frac{1}{2}}$.

A competing process along the edge is energy dissipation (loss to the environment). Heat, unlike charge, can escape from edge modes to phonons, photons (due to stray capacitances), or neutral excitations in the bulk (localized states coupled by Coulomb interaction). These processes cause an exponential decay in the upstream temperature, $T_u(L) = T_m e^{-\frac{L}{l_{\text{dis}}}}$, with $l_{\text{dis}}$ a characteristic dissipation length. Such dissipation is a compelling explanation to recent observations of relaxation of heat flow in particle-like states (with $n_u = 0$)[21]. If the thermal equilibration is weak compared to the energy dissipation ($l_{\text{dis}} \ll l_{\text{eq}}$), energy back-scattering is of no importance, and an exponentially decaying profile of $T_u$ is expected regardless of the state's topology[22].

In order to measure the local temperature of upstream modes, we fabricated devices based on a high mobility two-dimensional electron gas (2DEG), embedded in a GaAs-AlGaAs heterostructure, with density $8.2 \times 10^{10} cm^{-2}$ and mobility $4.4 \times 10^6 cm^2 V^{-1} s^{-1}$. Device A consists of three Ohmic contacts: source (S), an upstream located amplifier contact (A), and a ground contact (G) (see Fig. 1). The propagation length $L$, between S and A, could be varied using three metallic gates, which, when negatively charged, force the edge modes to take a detour, thus elongating the propagation length. When bias is applied to the source, power is dissipated at the back of the contact, leading to a hot spot[23] (depicted as a red fire) with an elevated temperature $T_m$.

The temperature of the upstream modes reaching A was determined from current fluctuations measured in A. This upstream noise is a smoking-gun signature of the presence of upstream modes[3,4], as studied theoretically recently[20,24]. The noise is generated in a *noise spot* (depicted as a white bolt sign), a region with size of the charge equilibration length outside A. The elevated temperature of upstream modes at this spot excites particle-hole pairs. If a particle (or hole) is absorbed by the amplifier contact while the hole (or particle) flows downstream, recombination does not take place and current fluctuations are detected in contact A. Thus, the local temperature of the upstream modes $T_u(L)$ is encoded in the excess noise

$$S_{\text{excess}}^{\text{U}} \propto (T_u(L) - T_0), \qquad (1)$$

where $T_0$ is the base temperature at A. The proportionality constant can depend on microscopic details of the edge modes, but importantly, it does not depend on $L$ (see Methods).

The measured noise is plotted in Fig. 2 at $\nu = \frac{2}{3}, \frac{3}{5}, 1+\frac{2}{3}, 1+\frac{3}{5}$, for $L$ in the range $30-210\mu m$. In Fig. 2a, the excess upstream noise is plotted as function of the voltage bias $V$ for three different $L$. Fig. 2b displays the noise measured at the fixed voltage $13\mu V$, plotted as a function of $L$ for several states from distinct topological classes: at $\nu = \frac{3}{5}$, $\nu_Q > 0$; at $\nu = \frac{2}{3}, 1+\frac{3}{5}$, $\nu_Q = 0$; and at $\nu = 1+\frac{2}{3}$, $\nu_Q > 0$. Remarkably, we find that all noise profiles are similar. The noise strength vs $L$ fits nicely to a decaying exponent with a characteristic decay length of $200\mu m$. This suggests an unequilibrated regime: $l_{\text{eq}} > l_{\text{dis}} \approx 200\mu m$.

Next, we studied the thermal conductance, which supplements the upstream thermometry, as it is sensitive only to the heat returning to the source contact (and not to dissipation). We used two devices, B1 and B2, where a floating Ohmic contact was employed as the heat source (Fig. 3a). The floating contact, with area of a few tens of $\mu m^2$, was connected to three separate arms of 2DEG. Simultaneously sourcing currents $+I$ and $-I$ from contacts $S_1$ and $S_2$ respectively leads to a dissipation of power $P = I^2/G_{2T}$ in the Ohmic contact, while its potential remains zero. Upstream and downstream noise was measured (Fig. 3b) at contacts (depicted as amp DS and amp US) located in the upstream and downstream direction. Similar to device A, the length between the Ohmic contact and the upstream amplifier contact could be varied. Devices B1 and B2 were designed for short $(15-85\mu m)$ and long $(35-315\mu m)$ upstream distances $L$ respectively.

The downstream current fluctuations further allow extraction of the Ohmic contact temperature $T_m$ (the downstream noise, unlike the upstream noise, is independent of the local temperature of upstream modes at the downstream amplifier), which in turn determines the thermal conductance via a heat balance equation (see Methods). The normalized thermal conductance vs $L$ is plotted in Fig. 3c. Remarkably, the thermal conductance is length-independent for both $\nu = \frac{2}{3}$ and $\nu = \frac{3}{5}$ up to lengths of $315\mu m$. On the other hand, the upstream noise decays by as much as 85% compared to the shortest length.

All measurements point to a lack of thermal equilibration, and instead to dissipation-dominated heat transport. As shown in Figs. 2b and 3c, the upstream noise decays exponentially with $L$ independently of $\nu$. If the edges were equilibrated, one would expect distinct behavior of upstream noise depending on the topological class due to different heat transport characteristics. Hence, the upstream-noise data shows that the thermal equilibration in the edge is not operative, and the noise decay is due to dissipation of energy to the environment. This is further supported by the length-independent thermal conductance at $=\frac{2}{3}$, which is incompatible with diffusive transport in the thermally equilibrated regime. Note that dissipation does not affect the thermal conductance in our measurement scheme since the dissipated heat does not return back to the Ohmic contact[13].

In the unequilibrated regime, the classical Johnson-Nyquist (JN) formula, used in similar experiments[10–12,25] should be corrected due to the mismatch between the upstream and downstream modes' temperature. The strength of the current fluctuations propagating downstream from a reservoir heated to a temperature $T_m$ is generally given by

$$S_{JN} = 2k_b G_{2T}(T_m - T_0)\alpha ,   \quad (2)$$

where $\alpha$ is a pre-factor that depends on microscopic details of the edge modes. The classical JN noise is restored ($\alpha = 1$) for a thermally equilibrated state with $n_d \geq n_u$, and in particular for any integer or particle-like fractional state (where $n_u = 0$). For unequilibrated $\nu = \frac{2}{3}$ and $\nu = \frac{3}{5}$ edges we find $\alpha = \frac{3}{4}$ and $\alpha = \frac{7}{10}$ respectively (see Methods).

At this point we can quantitively determine the thermal conductance. We plot the power $P$ dissipated at the Ohmic contact vs $T_m^2 - T_0^2$. A linear fit to the energy-balance equation $P = \frac{I^2}{G_{2T}} = 3\frac{\kappa_{2T}}{2}(T_m^2 - T_0^2)$ yields $\kappa_{2T}$ (see Methods). Note that $\kappa_{2T}$ determines the two-terminal thermal conductance of each individual arm, assuming that each arm contributes equally (which is the case in the absence of equilibration). For three integer states $\nu = 1, 2, 3$, the extracted thermal conductance agrees well with the expected values $\kappa_{2T}/\kappa_0 = n_d$ [see Supplementary Information (SI)]. Here, the absence of upstream modes in these states makes thermal equilibration irrelevant. For the hole-conjugate states $\nu = \frac{2}{3}$ and $\nu = \frac{3}{5}$, we find $\kappa_{2T}/\kappa_0 = 1.00 \pm 0.03$ and $\kappa_{2T}/\kappa_0 = 1.45 \pm 0.03$, respectively (see Fig. 4a). For completeness, we point out that if one derives $T_m$ with the classical JN formula ($\alpha = 1$)[10–12,25], a significantly higher thermal conductance value is obtained: $\kappa_{2T}/\kappa_0 = 1.5$ and $\kappa_{2T}/\kappa_0 = 2.5$ for $\nu = \frac{2}{3}$ and $\nu = \frac{3}{5}$ respectively.

How can we understand the measured values of $\kappa_{2T}$? As shown in Ref. [26] for $\nu = \frac{2}{3}$ [see SI for the generalization to $\nu = \frac{3}{5}$], an emerging quantization of $\kappa_{2T}$ is expected as

$$\kappa_{2T}/\kappa_0 = (n_u + n_d) - 2\kappa_{12}/\kappa_0, \qquad (3)$$

in an intermediate transport regime $L_T \ll L \ll l_{\text{eq}}$, where $L_T$ is the thermal length (see SI). The first term in Eq. (3) is the expected value in the absence of the thermal equilibration, but it is lowered by the parameter $2\kappa_{12}/\kappa_0$ due to backscattering of plasmon modes at boundaries between contacts and the edge. Generally, $\kappa_{12}$ depends on interactions on the edge. When the system is close to a low-energy fixed point[27–29] at which a charge mode is decoupled from neutral modes, we find $\kappa_{12}/\kappa_0 = 2(1-\nu)/(2-\nu)$. Then, $\kappa_{2T}/\kappa_0 = 1$ for $\nu = 2/3$ with $n_u = 1, n_d = 1$, and $\kappa_{2T}/\kappa_0 = 13/7$ for $\nu = 3/5$ with $n_u = 2, n_d = 1$. For $\nu = 2/3$, our measured value agrees very well with the prediction[26] $\kappa_{2T}/\kappa_0 = 1$. For $\nu = 3/5$, the measured value is slightly lower than $13/7$, which might be related to a deviation of the system from the infrared fixed point. Interestingly, for $\nu = 2/3$, it was recently calculated that at the fixed point, the thermal equilibration length diverges[25].

We turn now to a quantitative analysis of the upstream noise. With vanishing thermal equilibration, we can write:

$$S^{\text{U}}_{\text{excess}} = 2k_B G_{2T} f_T (T_m - T_0) \equiv 2k_B G_{2T} f_T \Delta T, \qquad (4)$$

with $T_u(L)$ replaced by $T_m$ in comparison to Eq. (1) (see SI). We denote the proportionality constant, $f_T$ as the *thermal Fano factor*. In Fig. 4b we plot the measured upstream excess noise as a function of $\Delta T$ for different lengths. All curves are linearly proportional to $\Delta T$, in agreement with Eq. (4). When $L$ decreases (and thus dissipation becomes less important), $f_T$ increases, but even at the shortest available length of $15\mu m$ it is ~2 times smaller than predicted by our microscopic calculations. The reason for this discrepancy remains to be understood.

In summary, we demonstrated that counter-propagating modes efficiently exchange charge, but not energy. While we estimate the charge equilibration length to be shorter than $5\mu m$ (see Methods), our observations set a lower bound on the thermal equilibration length $l_{\text{eq}} > l_{\text{dis}} \approx 200\mu m$. These observations seem to agree (at least qualitatively) with recent measurements in short (edge length ≈ $5\mu m$) graphene samples[25], but disagree with previous measurements in GaAs[10]. We also note that for co-propagating integer modes, the thermal equilibration has been reported to be much faster than charge equilibration[30]. Our observation of a thermally non-equilibrated transport regime for edges with counter-propagating modes provides important insights into the physics of hole-conjugate states and paves the way for the understanding of the exotic state at filling $\nu = 5/2$.

**Methods:**

**Sample preparation:**

The Ohmic contacts and gates were patterned using standard e-beam lithography-liftoff techniques. The Ohmic contact consists of Ni (7nm), Au(200nm), Ge(100nm), Ni (75nm), Au(150nm) alloyed at $440°C$ for 50 seconds. To minimize strain-induced reflection from the gates, the surface was covered by 7nm $HfO_2$ deposited at $200°C$. After deposition, we etched the $HfO_2$, (except under the gates) using *buffeed oxide etch*. The gate electrode consists of 5nm Ti and 15nm Au.

**Extraction of $T_m$ from downstream noise:**

The excess downstream noise $S_{\text{excess}}^{\text{D}}$ in the three-arm devices B1 and B2 (Fig. 3a) is given by

$$S_{\text{excess}}^{\text{D}} = \frac{2}{3}\overline{(\Delta I_m)^2} + \frac{1}{9}\left(S_{\text{excess}}^{S_1} + S_{\text{excess}}^{S_2} + S_{\text{excess}}^{\text{U}}\right) - \frac{4}{3}G_{2T}k_B T_0, \quad \text{(M1)}$$

where the numerical factors come from the devices' geometry (see SI for details). Here, $\overline{(\Delta I_m)^2}$ is the noise from the current fluctuations emanating from the central floating contact and are thus related to the temperature $T_m$ of the central contact, while $S_{\text{excess}}^{S_1}$ and $S_{\text{excess}}^{S_2}$ are the excess noises generated at noise spots (SI, Fig. S6) near sources $S_1$ and $S_2$. Since $S_{\text{excess}}^{S_1}$ and $S_{\text{excess}}^{S_2}$ are analogous to the upstream noise, they can be taken into account by using measurements of $S_{\text{excess}}^{\text{U}}$ at upstream distances $30\mu m$ (corresponding to the distance between $S_1$ and the central contact) and $150\ \mu m$ (the distance between $S_2$ and the central contact), respectively.

The generated noise $\overline{(\Delta I_m)^2}$ on an edge segment is generally given by[20]

$$\overline{(\Delta I_m)^2} = \frac{2e^2}{h l_{\text{eq}}^C}\frac{\nu_-}{\nu_+}(\nu_+ - \nu_-)\int_0^L dx\, \Lambda(x) e^{-2x/l_{\text{eq}}^C} + \frac{2e^2}{h}k_B T_m \frac{(\nu_+ - \nu_-)^2}{\nu_+}, \quad \text{(M2)}$$

where $l_{\text{eq}}^C$ is the charge equilibration length and $\nu_+(\nu_-)$ is the total filling factor of the downstream (upstream) modes (e.g., $\nu_+ = 1$, $\nu_- = 1/3$ for $\nu = 2/3$ and $\nu_+ = 1$, $\nu_- = 2/5$ for $\nu = 3/5$ ). The noise $\overline{(\Delta I_m)^2}$ is the local noise generated by inter-channel electron tunneling, which is encoded in the noise kernel $\Lambda(x)$. Assuming absence of thermal equilibration, $\Lambda(x)$ becomes independent of position $x$ [i.e., $\Lambda(x) = \Lambda_0(T_m, T_0)$], and thus Eq. (M2) is simplified as

$$\overline{\Delta I_m^2} = \frac{e^2}{h}\frac{\nu_-}{\nu_+}(\nu_+ - \nu_-)\Lambda_0(T_m, T_0) + \frac{2e^2}{h}k_B T_m \frac{(\nu_+ - \nu_-)^2}{\nu_+}. \tag{M3}$$

The simplified noise kernel $\Lambda_0(T_m, T_0)$ can be computed within a microscopic model (see SI for details). Specifically, we divide the edge segment into three regions: the left contact region, a central region, and the right contact region. The inter-channel interaction is taken to change sharply from zero in the contact regions to a finite value in the central region. The left and right contacts are taken at different temperatures $T_m$ and $T_0$ respectively. Within this model, we derive the formula $\Lambda_0(T_m, T_0) \simeq 2T_0 + 0.5\,(T_m - T_0)$, assuming strong interactions. The downstream excess noise $S^D_{\text{excess}}$ then reads

$$S^D_{\text{excess}} = \frac{1}{9}\left(S^{S_1}_{\text{excess}} + S^{S_2}_{\text{excess}} + S^U_{\text{excess}}\right) + \frac{4}{3}G_{2T}k_B(T_m - T_0)\frac{4\nu_+ - 3\nu_-}{4\nu_+}. \tag{M4}$$

Comparing to Eq. (2) and identifying $S_{JN} = \overline{(\Delta I_m)^2} - 2k_B G_{2T} T_0$, we find $\alpha = \frac{4\nu_+ - 3\nu_-}{4\nu_+}$. The central Ohmic contact temperature $T_m$ can be extracted from Eq. (M4). Ultimately, $T_m$ is used for the determination of both the thermal conductance $\kappa_{2T}$ and in plotting the upstream noise vs $\Delta T = T_m - T_0$.

**Extraction of $\kappa_{2T}$:**

The thermal conductance $\kappa_{2T}$ is obtained from the heat balance equation

$$P = \frac{I^2}{G_{2T}} = \frac{3}{2}\kappa_{2T}(T_m^2 - T_0^2), \tag{M5}$$

which, in the steady state, equates the dissipated power and the emanating heat currents in the central contact. Here, we assume that all injected electrical power is dissipated in the central contact and raises its temperature (see SI for more details). The temperature $T_m$ in Eq. (M5) is extracted from downstream noise, Eq. (M4) as described above.

**Upstream noise theory:**

With the same model as for the downstream noise, the upstream excess noise is computed as

$$S^U_{\text{excess}} = \frac{3}{2}\frac{e^2}{h}\frac{\nu_-}{\nu_+}(\nu_+ - \nu_-)(T_m - T_0). \tag{M6}$$

The derivation of Eq. (M6) is given in the SI. In comparison to Eq. (4), we have $f_T = \frac{3}{4}\frac{\nu_-}{\nu_+}$. Equation (M6) is used for the theoretical plots in Fig. 4b.

**Estimation of charge equilibration length:**

A slightly unequilibrated charge conductance was reported at $\nu = \frac{2}{3}$ for very short edge distances[31]. In order to test the equilibration of charge we sourced from $S_1$ AC voltage at the resonance frequency of the upstream amplifier (not DC current like the main measurements) and measured the voltage in the upstream amplifier. Given the sourced voltage $V_s$, we find by using the standard Landauer-Büttiker formalism,

$$V_{\text{amp}} = V_S \frac{G_U(L)}{3G_{2T}}, \tag{M7}$$

where $G_U(L)$ is the upstream conductance from the floating Ohmic contact to the upstream amplifier (the factor of 3 comes from the three arms of the B1 and B2 devices). In deriving Eq. (M7), we assumed $G_U \ll G_{2T}$. We see (SI, Fig. S5) that only at $\nu = \frac{2}{3}$ and $= \frac{3}{5}$, and only when the propagation length is short, one would observe a finite $G_U$, which would indicate a not fully charge equilibrated edge. At the shortest available length $L = 15\mu m$, we found $\frac{G_U}{G_{2T}} = 7 \times 10^{-3}$ for $\nu = \frac{2}{3}$ and $\frac{G_U}{G_{2T}} = 3 \times 10^{-4}$ for $\nu = \frac{3}{5}$. To rule out the possibility that the upstream current is a result of bulk currents due to finite longitudinal conductance, we repeated the measurement at a higher temperature. We observed that $G_U$ decreases at $21 mK$, as apparently the charge equilibration is faster. This behavior is inconsistent with bulk currents, since the longitudinal conductance is expected to increase with temperature. In a simplle model for charge equilibration, we can write

$$\frac{G_U(L)}{G_{2T}} = G_{U,0} e^{-\frac{L}{l_{\text{eq}}^C}}, \tag{M8}$$

where $G_{U,0} = \frac{e^2}{h}\nu_-$ is the zero length upstream conductance. From our data, we find $l_{\text{eq}}^C \approx 4\ \mu m$ for $\nu = \frac{2}{3}$ and $l_{\text{eq}}^C \approx 2\ \mu m$ for $\nu = \frac{3}{5}$. This charge equilibration length stands in sharp contrast to our observed thermal equilibration lengths which are two orders of magnitude larger.

**Acknowledgments**

We thank M. Heiblum for his essential guidance and support throughout this project. We also thank Y. Gefen for many stimulating discussions in the course of this work. We acknowledge the help of Diana Mahalu with e-beam lithography. We acknowledge illuminating discussions with K. Snizhko.

C.S. acknowledges funding from the Excellence Initiative Nano at Chalmers University of Technology. J.P. acknowledges funding by the Deutsche Forschungsgemeinschaft (DFG, German Research Foundation) – Projektnummer 277101999 – TRR 183 (project A01). C.S. and A.D.M. acknowledge support by DFG Grants No. MI 658/10-1 and MI 658/10-2, and by the German-Israeli Foundation Grant No. I-1505-303.10/2019.

**Author contributions**

R.A.M and B.D. fabricated the devices and performed the measurements. R.A.M analyzed the data with inputs from B.D, C.S, J.P, and A.D.M. C.S, J.P, and A.D.M developed the theoretical model. V.U grew the GaAs heterostructures. All authors contributed to the writing of the manuscript.

# Figures:

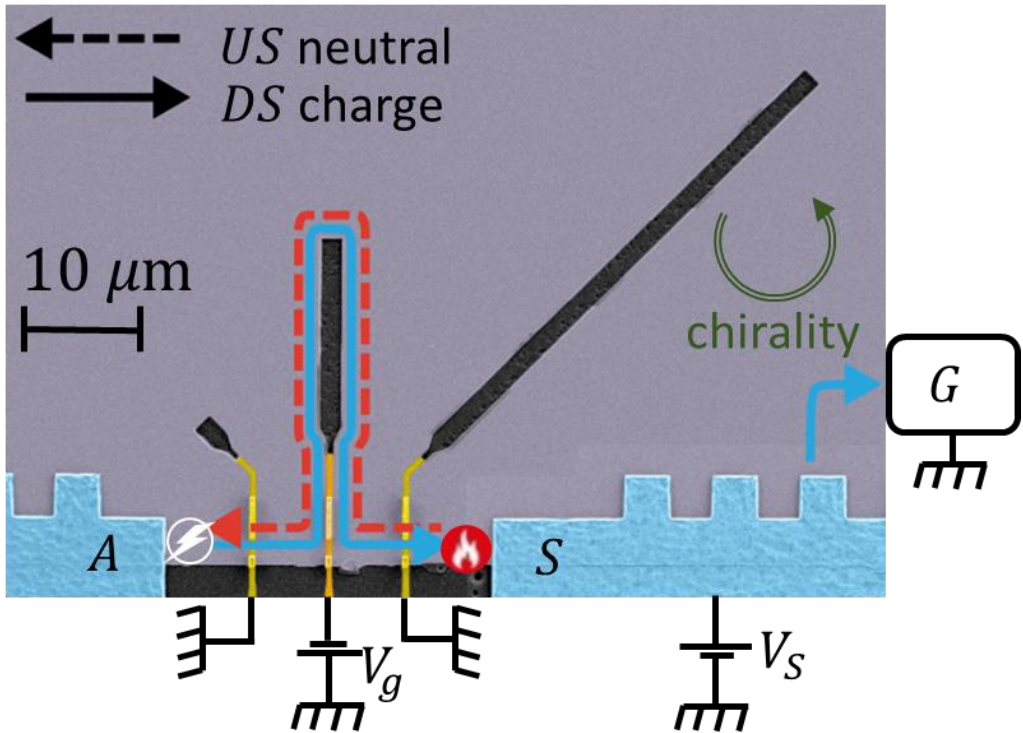

**Figure 1 | Device.** False colors SEM image of the heart of Device A. This device consists of three Ohmic contacts at the edge of the MESA (colored grey): Source(S), Amplifier(A) and the cold-grounded drain (G) (shown by symbol only). The propagation length of the edge modes between the S and A contacts can be tuned by using the three metallic gates (light-yellow: unbiased, dark-yellow: biased), which upon biasing, add the etched regions inside the MESA (funnel shaped black regions) to the upstream path. Applying a voltage $V_s$ on S causes the formation of a hot-spot (marked with red fire) at the back of S. The upstream modes (red dashed line) emanating from the hot-spot, carry the heat to the A where the noise is generated (marked with white bolt). We depict here the path of the edge modes for the situation where the middle gate (dark yellow) redirects the edge modes with the application of a gate voltage $V_g$, while the other two gates remain unbiased (light yellow), and hence do not affect the edge modes' path. Zero bias on all gates forms the shortest propagation length (straight line from S to A), while biasing all three gates forms the longest propagation length.

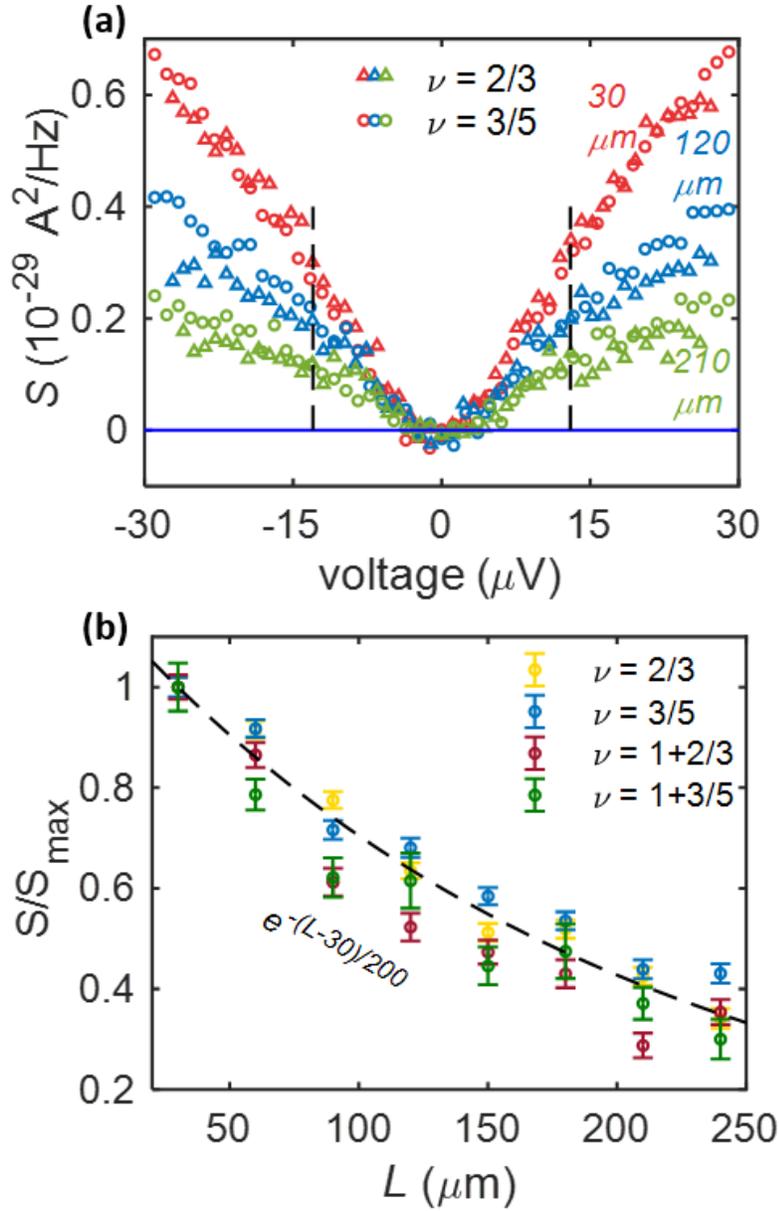

**Figure 2 | Length Profile of the upstream noise in different states. a,** Upstream noise as a function of the applied bias to the Source for $\nu = \frac{2}{3}$ (triangles) and $\nu = \frac{3}{5}$ (circles), for a few propagation lengths; $30\mu m$ (red), $120\mu m$ (blue), $210\mu m$ (green). The dashed lines mark the voltage for which the length dependence profile was determined. **b,** Length dependence of the upstream noise. The noise is normalized (with respect to shortest length) separately for each filling factor. The noise profile of all fillings matches nicely with exponential decay with a typical decay length of $200\mu m$ (dashed line). This indicates the dominant role of dissipation rather than thermal equilibration.

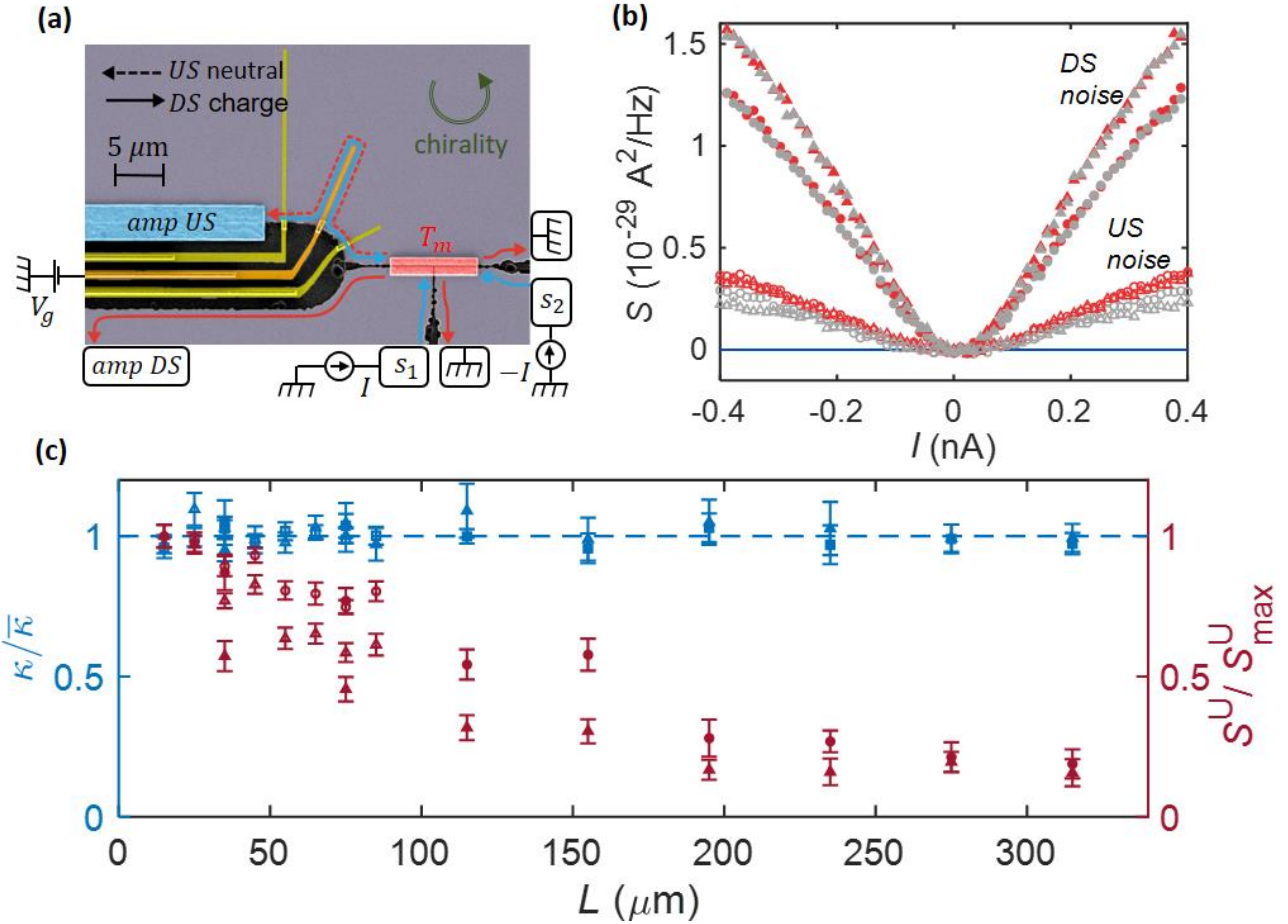

**Figure 3 | Length profile of the thermal conductance and the upstream noise. a,** False colors SEM image of the central part of Device B1. The mesa (grey) is divided into three arms by the etched regions (black). The three arms are connected by a floating metallic island (in red) with area 15 x 2 µm², serving as a heat source. When a current $I$ from $S_1$ and $-I$ from $S_2$ are sourced simultaneously, the floating island heats up to a temperature $T_m$. The resulting noise is measured simultaneously in the DS and US amplifiers. The propagation length from the floating contact to the US amplifier can be varied using the metallic gates (yellow, as in Device A). Depicted is the case where the middle gate (darker yellow) redirects the path of the edge modes by the application of a gate voltage $V_g$, while the other gates are unbiased, and hence do not affect the propagation length. **b,** DS noise (full shapes) and US noise (empty shapes) as a function of the current. Results are shown for $\nu = \frac{2}{3}$ (triangles) and $\nu = \frac{3}{5}$ (circles), and the propagation lengths $15\mu m$ (red) and $75\mu m$ (grey). The US noise decays with length while the DS noise does not. **c,** Two terminal thermal conductance $\kappa_{2T}$ (extracted from the DS noise) (blue), and US noise strength (red) as a function of length (See Methods). The thermal conductance is separately normalized for $\nu = \frac{2}{3}$ and $\nu = \frac{3}{5}$ with respect to their respective means. For both $\nu = \frac{2}{3}$ (triangles) and $\nu = \frac{3}{5}$ (circles), we observe that $\kappa_{2T}$ is length independent, while the US noise decays (similarly to Fig. 2b). This indicates an unequilibrated thermal regime. The empty (full) shapes mark the data measured in device B1 (B2).

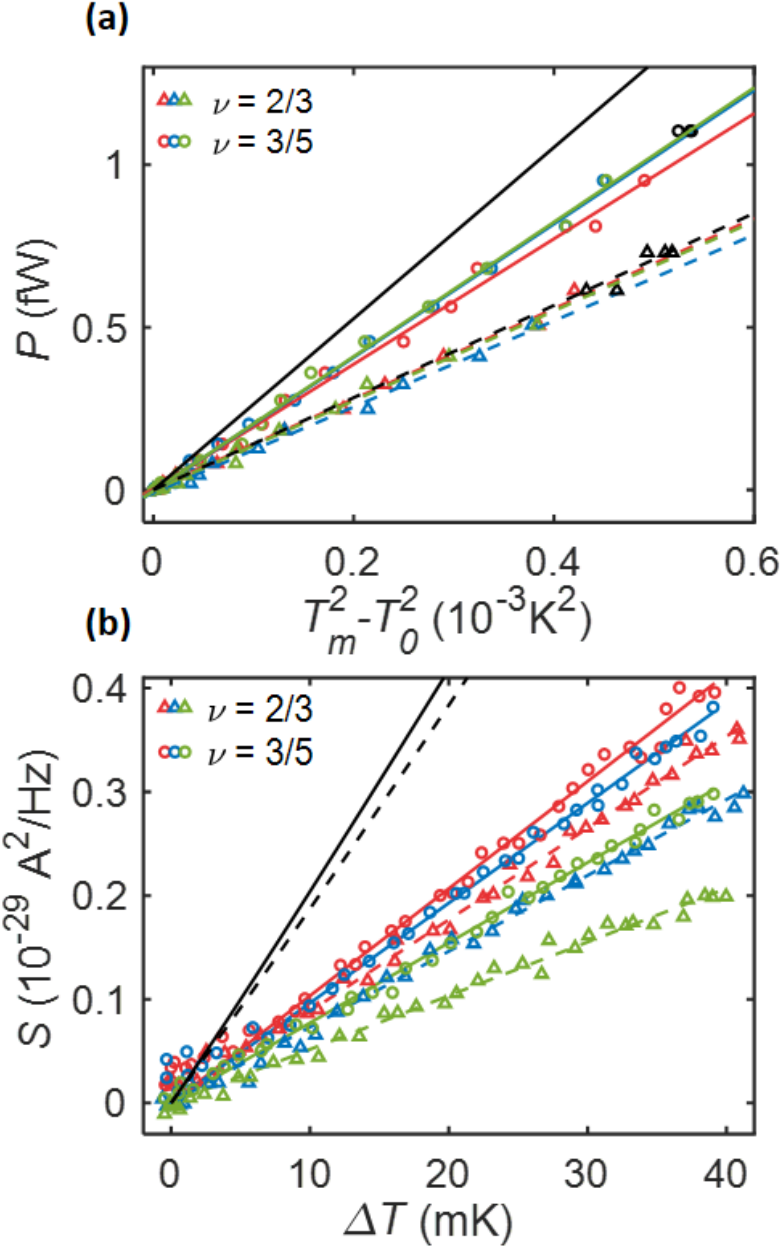

**Figure 4 | Quantitative analysis of the thermal conductance and the upstream noise. a,** Dissipated power P as a function of $T_m^2 - T_0^2$, where $T_m$ and $T_0$ are the Ohmic contact and the base temperatures respectively. The colored markers (low temperature data - $T_m < 25mK$) were linearly fitted to extract $\kappa_{2T}$ (fits marked by colored dashed and full lines for $\nu = \frac{2}{3}$ and $\nu = \frac{3}{5}$ respectively). The black markers are high temperature points and were not fitted. We plot the data for $\nu = \frac{2}{3}$ and $\nu = \frac{3}{5}$ for propagation lengths $15\mu m$ (red) $45\mu m$ (blue) and $85\mu m$ (green). We find length independent, thermal conductances $\kappa_{2T}/\kappa_0 = 1.00 \pm 0.03\kappa_0$, $\kappa_{2T}/\kappa_0 = 1.45 \pm 0.03\kappa_0$ for $\nu = \frac{2}{3}$ and $\nu = \frac{3}{5}$ respectively. The theoretically expected values for $\nu = \frac{2}{3}$ ($\nu = \frac{3}{5}$) are plotted as a black dashed (full) line. We find excellent agreement with the data for $\nu = \frac{2}{3}$, while the thermal conductance for $\nu = \frac{3}{5}$ is somewhat smaller than predicted. **b,**

Excess US noise as a function of $T_m$ for $\nu = \frac{2}{3}$ and $\nu = \frac{3}{5}$, and for propagation lengths $15\mu m$ (red), $45\mu m$ (blue) and $85\mu m$ (green). The slope of the linear fit, denoted as $2k_B G_{2T} f_T$ in Eq. (3), increases with decreasing length (due to diminishing dissipation) and approaches a value of roughly 0.5 times that predicted by a microscopic calculation. The predicted values (see SI) are depicted by the black, dashed and solid line for $\nu = \frac{2}{3}$ and $\nu = \frac{3}{5}$ respectively.

# Supplementary Information for "Absent thermal equilibration on fractional quantum Hall edges over macroscopic scale"

Ron Aharon Melcer,[1] Bivas Dutta,[1] Christian Spånslätt,[2,3,4] Jinhong Park,[5] Alexander D. Mirlin,[3,4,6,7] and Vladimir Umansky[1]

[1]*Braun Center for Submicron Research, Department of Condensed Matter Physics, Weizmann Institute of Science, Rehovot 761001, Israel*

[2]*Department of Microtechnology and Nanoscience (MC2), Chalmers University of Technology, S-412 96 Göteborg, Sweden*

[3]*Institute for Quantum Materials and Technologies, Karlsruhe Institute of Technology, 76021 Karlsruhe, Germany*

[4]*Institut für Theorie der Kondensierten Materie, Karlsruhe Institute of Technology, 76128 Karlsruhe, Germany*

[5]*Institute for Theoretical Physics, University of Cologne, Zülpicher Str. 77, 50937 Köln, Germany*

[6]*Petersburg Nuclear Physics Institute, 188300 St. Petersburg, Russia*

[7]*L. D. Landau Institute for Theoretical Physics RAS, 119334 Moscow, Russia*

(Dated: June 23, 2021)

# S1. TECHNICAL DETAILS OF NOISE MEASUREMENTS

As described in the main text, upstream and downstream noise was measured in two distinct amplifiers. Both amplifiers are connected in parallel to an RLC circuit. In our case the inductance comes from a superconducting coil located at the mixing chamber plate, the capacitance is the line capacitance leading from the sample to the homemade cryo-amplifier located at the $4.2K$ plate, and $R = G_{2T}^{-1} = h/(\nu e^2)$ is the sample resistance. We configured the inductance of the coils such that the resonance curves of the two amplifiers don't overlap: the central frequencies are $f_D = 693$kHz and $f_U = 633$kHz (with indices D and U denoting "downstream" and "upstream", respectively). The band width of both circuits is 15kHz (14kHz) for $\nu = 2/3$ ($\nu = 3/5$). In this case, in the frequency range being picked up by the downstream amplifier, the upstream amplifier contact is simply a short to ground (see Fig. S1 for a detailed description of the measurement circuit).

This configuration simplifies the analysis of the downstream noise, which is used for extraction of the temperature $T_m$ of the central floating island $\Omega_m$. This is described in more detail in Sec. S12. The measured noise in the downstream amplifier includes several contributions. The major contribution comes from Johnson-Nyquist noise generated at the central floating contact due to its elevated temperature. A second contribution comes from reflected upstream noise. In the frequency bandwidth where the downstream noise is measured, both the upstream amplifier contact and the source contacts $S_1$ and $S_2$ are effectively grounded (the source contacts are connected via a 5nF capacitance to ground). Therefore, the current noise generated at the noise spots, next to these contacts, flows to $\Omega_m$, and from there to the downstream amplifier. For more details, see Sec. S6 B 1 below.

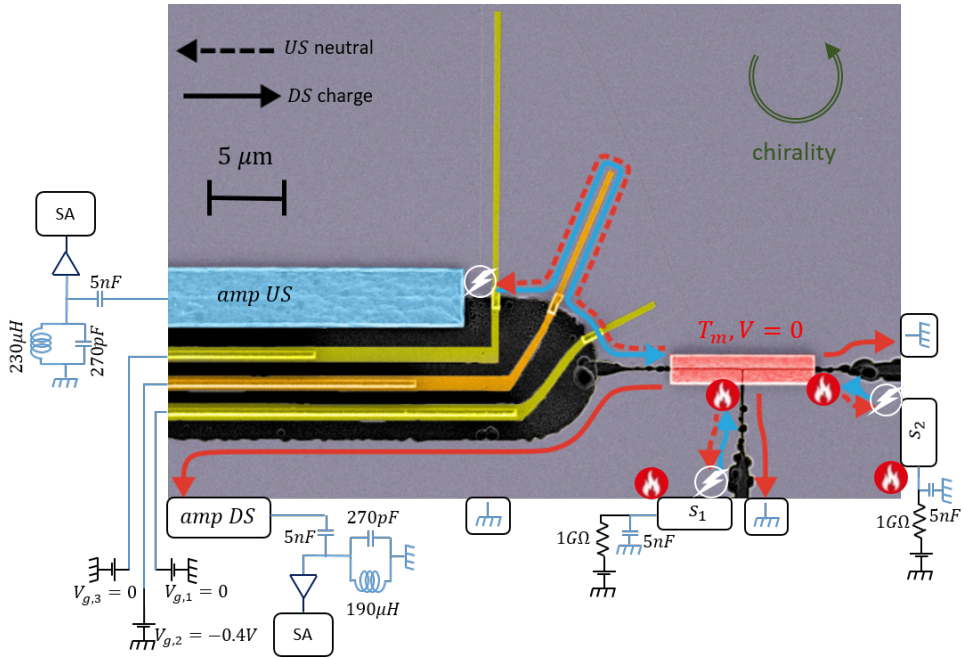

FIG. S1. False color SEM image of device B1, similar to Fig. 3a in the main text. Here, we have denoted all hot spots and noise spots that form in the experiment. In addition, we have specified the circuit components used in our measurements: the color of a component corresponds to its temperature: room temperature (black), 4.2K (dark blue) and 6mK (light blue).

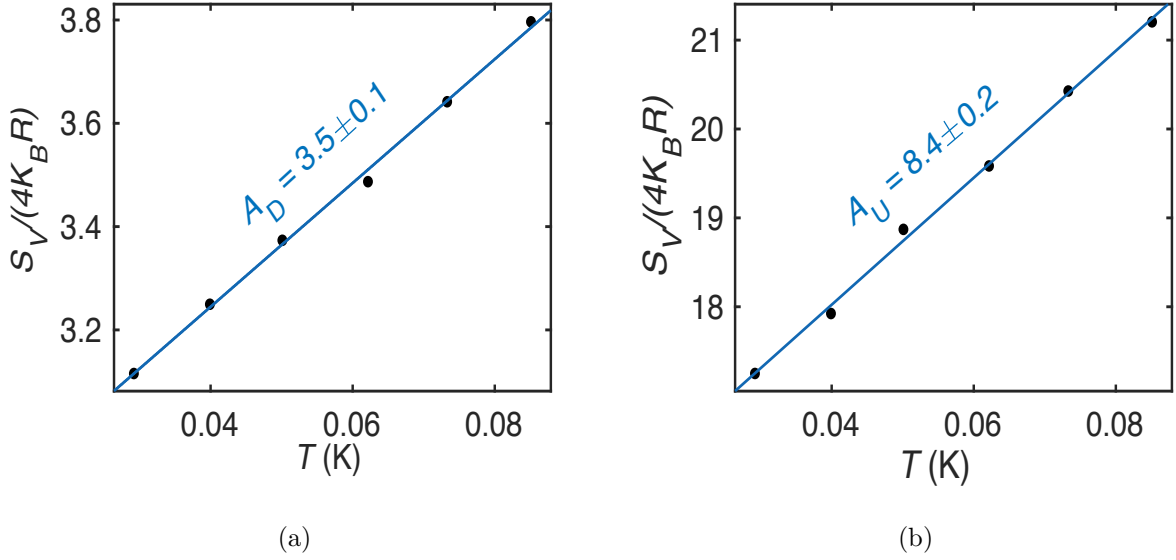

FIG. S2. Amplifier calibration. Voltage noise strength $S_V$ as a function of the mixing chamber temperature $T$ for (a) the downstream amplifier and (b) the upstream amplifier. The slope of the linear fit is the gain squared ($A^2$).

## S2. AMPLIFIER GAIN AND TEMPERATURE CALIBRATION

The method used to calibrate the amplifier gain and electron temperature is based on Johnson-Nyquist noise[S1,S2]: $S_V = 4A^2 k_B T / G_{2T}$, where $A$ is the gain of the amplifiers, $k_B$ is the Boltzmann constant, and $T$ is the temperature. Linearly fitting the noise with the fridge temperature (see Fig. S2), allows us to extract the gain $A$ and the base noise of the amplifiers $S_{\text{base}}$. The base noise is used to extract the temperature $T_0$ of the electrons. At the lowest temperature (typically below 20mK) the electron temperature can be higher than that of the cryostat. We extract $T_0$ from: $T_0 = (S_V - S_{\text{base}})G_{2T}/(4A^2 k_B)$. In our measurements the cryostat temperature was always 6mK and $T_0$ was measured between 11mK and 14mK for all considered Hall states. An interesting observation is that $T_0$ tends to be smaller for states with larger $\kappa_{2T}$, such as $\nu = 3$ and $\nu = 3/5$, and a somewhat larger for states with small $\kappa_{2T}$ such as $\nu = 1$ and $\nu = 2/3$. This observation suggests that at low temperatures, the most efficient cooling mechanism of the devices is via the edge modes.

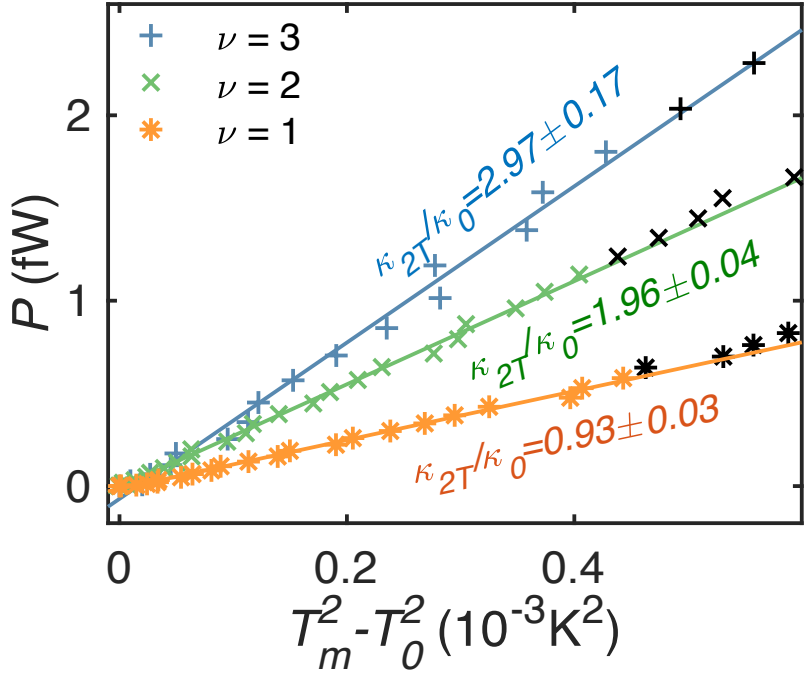

FIG. S3. Injected power $P$ vs $T_m^2 - T_0^2$, where $T_m$ and $T_0$ are the Ohmic contact and base temperatures respectively. The colored markers (low temperature data $T_m < 25$mK) were linearly fitted to extract $\kappa_{2T}$. The black markers are for high temperature points, which were not included in the fit. Data is plotted for $\nu = 1, 2, 3$ in orange, green, and blue respectively. The slopes of the linear lines fit well to expected values of $\kappa_{2T}/\kappa_0 = n_d = 1, 2, 3$ respectively.

## S3. THERMAL CONDUCTANCE MEASUREMENT OF INTEGER STATES

To check that the thermal conductance measurements work properly in our $B$-devices (depicted in Fig. S1), we performed thermal conductance measurements at the three integer fillings $\nu = 1, 2, 3$ (see Fig. S3). The measurement scheme was identical to the fractional states and is described in the main text (though for the integer states, all length-controlling gates were not operational, thereby fixing $L$ to the smallest available value). We find good agreement with the expected quantizations $\kappa_{2T}/\kappa_0 = n_d$. For the integer states, where there are only downstream modes, neither equilibration nor dissipation influences the thermal conductance. The suppression of a single quantum of thermal conductance due to heat Coulomb blockade[S3,S4] was not observed in our devices due to the relatively large central Ohmic contact.

## S4. ESTIMATES OF THE THERMAL EQUILIBRATION AND DISSIPATION LENGTHS

In the main text, we demonstrated that the decay of the upstream noise with distance $L$ is essentially the same for all considered filling fractions, and is well described by a simple exponential with the decay length $\approx 200\,\mu$m (see Fig. 2b in the main text). This observation strongly suggests that the dominant mechanism determining the decay of the upstream noise is dissipation (leakage) of the energy from the quantum-Hall edge to the environment, with the characteristic decay length $l_\text{dis} \approx 200\,\mu$m. At the same time, inter-mode thermal equilibration is not operative on the studied distances. In this section, we present further details of the analysis of the data and the fitting procedure supporting these conclusions.

In Ref. S5, the authors derived, using a phenomenological model, a generic formula for the temperature of upstream modes, $T_U(L)$ in the presence of equilibration and dissipation. When the upstream modes are sourced at a temperature $T_m$ and the distance is $L$, then

$$T_U^2(L) = T_0^2 + \frac{1}{2} \frac{\Lambda(T_m^2 - T_0^2)}{(N/(2\tilde{n}) + l_\text{eq}/l_\text{dis})\sinh[\Lambda L/l_\text{eq}] + (\Lambda/2)\cosh[\Lambda L/l_\text{eq}]} e^{-\frac{L}{\tilde{n}l_\text{eq}}}. \tag{S1}$$

Here, $\tilde{n} = (n_u n_d)/(n_d - n_u)$, $N = (n_u + n_d)/(n_d - n_u)$, and $\Lambda = \sqrt{\frac{1}{\tilde{n}^2} + 4\frac{l_\text{eq}}{l_\text{dis}}\left(\frac{N}{\tilde{n}} + \frac{l_\text{eq}}{l_\text{dis}}\right)}$. In order to have a quantitative bound on $l_\text{eq}$, we fit our measured results from device A (Fig. 2b in the main text) to Eq. (S1). We assume that both $l_\text{eq}$ and $l_\text{dis}$ do not change between different states and that these parameters are the approximately the same for different modes on the same edge. We calculate the goodness of the fit of the model described by Eq. (S1) with different equilibration and dissipation lengths. The only parameter that we fit separately for each state is the noise amplitude at the shortest length (which is affected by micropscopic properties of the hot-spot, and thereby beyond experimental control). The

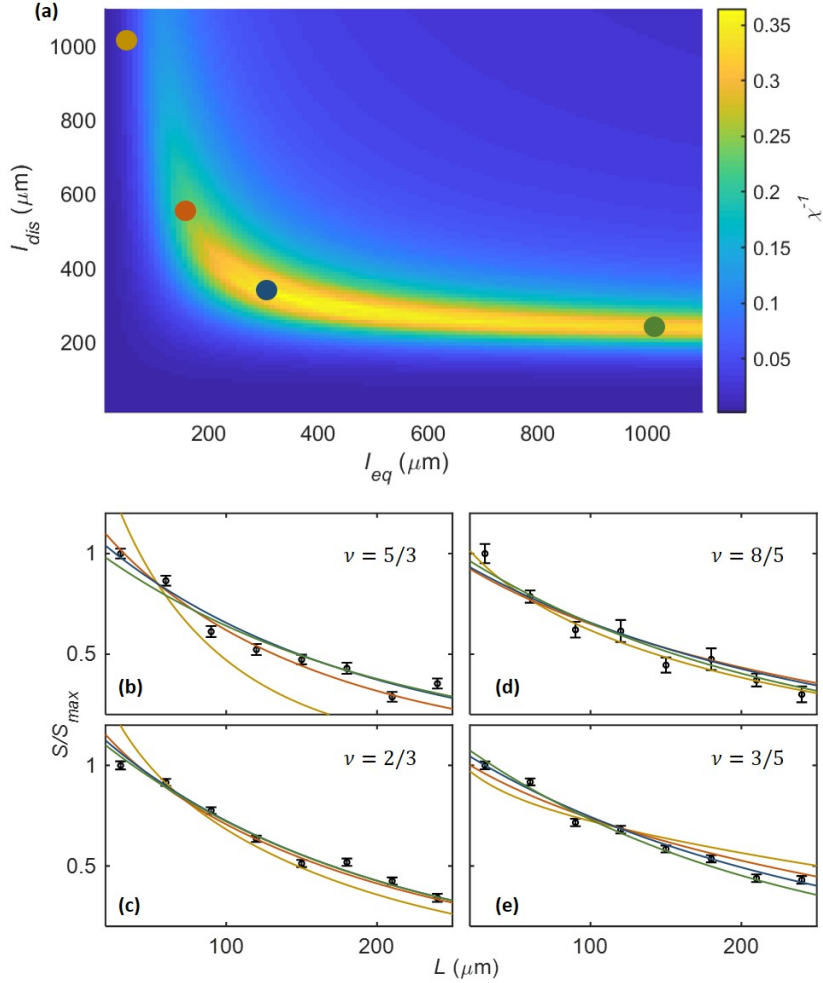

FIG. S4. (a) Goodness of fits for the measured noise profiles, as a function of the equilibration length $l_{eq}$ and the dissipation length $l_{dis}$. No good fit exists for $l_{eq} < 200 \mu m$. On the other hand, our data is consistent with $l_{eq} \to \infty$ and $l_{dis} \approx 200 \mu m$. The four colored points are at: $l_{eq} = 1050 \mu m$, $l_{dis} = 220 \mu m$ (green); $l_{eq} = 330 \mu m$, $l_{dis} = 330 \mu m$ (blue); $l_{eq} = 150 \mu m$, $l_{dis} = 550 \mu m$ (orange); $l_{eq} = 50 \mu m$, $l_{dis} = 1050 \mu m$ (yellow), and correspond to the four fits presented in (b-e). While the green and blue curves fits well to all filling factors, the orange curve fits poorly at $\nu = 3/5$, and the yellow curve fits poorly at $\nu = 5/3$ and $\nu = 3/5$.

goodness of the fit is defined as

$$\chi^2(l_{eq}, l_{dis}) = \frac{1}{N} \sum \frac{(S(l_{eq}, l_{dis}, L) - S^U_{excess}(L))^2}{\sigma(L)^2}, \tag{S2}$$

where $S(l_{eq}, l_{dis}, L) \propto T_U(L)$ is the expected noise according to Eq. (S1), $S^U_{excess}(L)$ is the measured noise, $\sigma(L)$ is the uncertainty in the noise, and $N = 32$ is the total number of measured points (eight points for each of four different filling factors). The sum is over all the different lengths of all the different fillings. In Fig. S4(a) we show $\chi^{-1} = (\overline{\chi^2})^{-1/2}$ as a function of $l_{\text{eq}}$ and $l_{\text{dis}}$. We see that our measurements are consistent with a very long equilibration length, and a dissipation length of roughly $200\mu m$. The goodness of the fit becomes poor for $l_{\text{eq}} < 200\mu m$, for any value of $l_{\text{dis}}$. The corresponding fits at four distinct points are presented in Fig. S4(b-e) for all the measured states.

## S5.  ESTIMATION OF THE CHARGE EQUILIBRATION LENGTH

It was predicted in Ref. S6 that the two-terminal electric conductance exhibits a crossover from the non-equilibrated to equilibrated value (e.g., from $4/3$ to $2/3$ at $\nu = 2/3$) when the length is increased. Such a crossover was subsequently observed experimentally in an engineered $\nu = 2/3$ edge[S7]. In contrast, almost all previous experiments on conventional edges exhibited equilibrated values of the electric two-terminal conductance (i.e., $G_{2T} = \nu e^2/h$), indicating very short charge equilibration lengths. A slight deviation in the conductance value indicating incomplete charge equilibration has been reported at the non-engineered $\nu = 2/3$ edge for very short edge distances[S8]. In order to study the equilibration of charge in our devices, we sourced an AC voltage at the resonance frequency of the upstream amplifier (i.e., not a DC current like in the main measurements) from $S_1$ and measured the resulting voltage in the upstream amplifier. Given a source voltage $V_S$, and assuming that $G_U \ll G_D$, we get using standard Landauer-Büttiker formalism

$$V_{\text{amp}} = V_S \frac{G_U}{3G_{2T}}. \tag{S3}$$

Here, $G_U$ and $G_D$ are the conductances from $\Omega_m$ to the upstream (downstream) amplifier, with $G_U + G_D = G_{2T}$. The factor of 3 in Eq. (S3) comes from the three arms of the device. For full charge equilibration, the entire charge current flows downstream. Thus, the ratio $G_U/G_{2T}$ serves as a quantitative measure of deviation from full charge equilibration.

Our results for $G_U/G_{2T}$ are shown in Fig. S5. For $\nu = 1$ and $\nu = 3$, we find $G_U/G_{2T} = 0$ as expected: for integer filling factors, the edge hosts only downstream modes, i.e. there is only downstream charge transport. By contrast, for $\nu = 2/3$ and $\nu = 3/5$, and for short propagation lengths, we observe non-zero values $G_U/G_{2T}$, indicating incomplete charge equilibration. However, the observed values of $G_U/G_{2T}$ are very small even for the shortest length of 15$\mu$m. Specifically, we found $G_U/G_{2T} = 7 \times 10^{-3}$ and $3 \times 10^{-4}$ for $\nu = 2/3$ and $\nu = 3/5$, respectively. This implies that these edges are nearly fully electrically equilibrated already for this short length, i.e., the charge equilibration length is substantially shorter than 15$\mu$m.

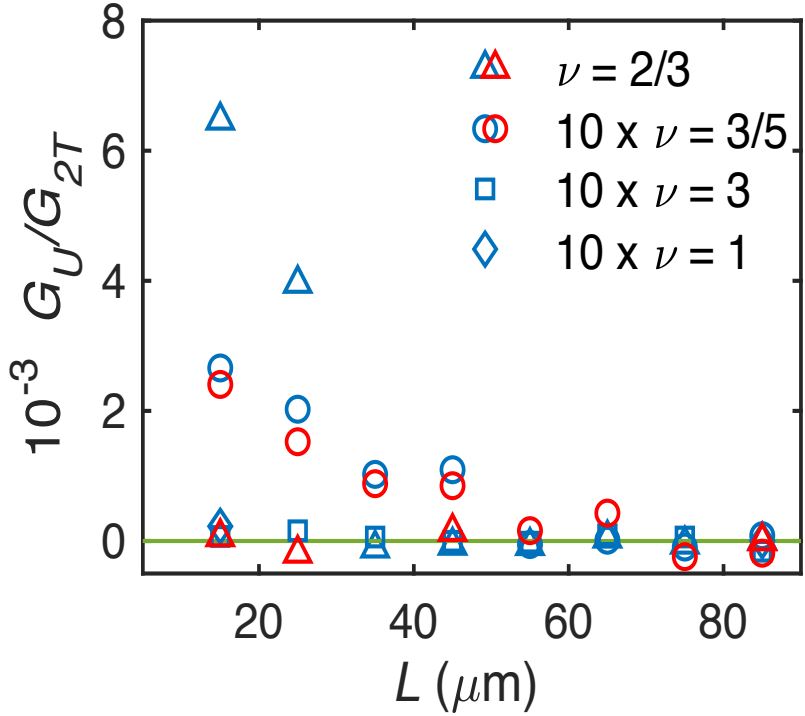

FIG. S5. Upstream charge conductance $G_U$ (normalized to $G_{2T} = \nu e^2/h$) vs propagation length $L$. When $L$ is short, there is a finite upstream conductance for $\nu = 2/3$ and $\nu = 3/5$, but not for $\nu = 3$ and $\nu = 1$. The upstream conductance is always small (below $7 \times 10^{-3} G_{2T}$ for $\nu = 2/3$ and below $3 \times 10^{-4} G_{2T}$ for $\nu = 3/5$) and decays with $L$, in accordance with charge equilibration. At a higher temperature of $T_0 = 21$mK (red markers), $G_U$ decreases (vanishes fully for $\nu = 2/3$), ruling out the possibility that the upstream current is a result of a finite longitudinal conductance. Blue markers denote a lower temperature $T = 11$mK for $\nu = 3$ and $\nu = 3/5$, $T = 13$mK for $\nu = 1$, $T = 14$mK for $\nu = 2/3$, respectively.

In order to rule out the possibility that the upstream current is a result of bulk currents due to finite longitudinal conductivity, we repeated the measurement at a higher temperature. We observed that $G_U$ decreases when the temperature is raised to 21mK. This behavior is consistent with charge equilibration, since the charge equilibration length is expected to in-crease with increasing temperature. At the same time, upstream transport via bulk currents would show an opposite behavior, as the longitudinal conductivity is expected to increase with temperature. Thus, the observed temperature dependence confirms that the non-zero values of $G_U$ are due to incomplete electric equilibration between the counterpropagating modes. To estimate the value of the corresponding equilibration length $l^C_{\text{eq}}$, we recall[S6,S9] that the conductance approaches exponentially its limiting (non-equilibrated) value for $L \gg l^C_{\text{eq}}$. Thus, we estimate the $L$ dependence of $G_U$ for $\nu = 2/3$ and $\nu = 3/5$ via

$$G_U(L) = G_U(0)\, e^{-L/l^C_{\text{eq}}}, \tag{S4}$$

where $G_U(0) = e^2 \nu_-/h$ is the zero-length upstream conductance, and $\nu_-$ is the total filling factor of the upstream modes. We find $l^C_{\text{eq}} \approx 4\mu$m for $\nu = 2/3$ (with $\nu_- = 1/3$) and $l^C_{\text{eq}} \approx 2\mu$m for $\nu = 3/5$ ($\nu_- = 2/5$). These small values of the charge equilibration lengths stand in sharp contrast to the much larger estimate of the thermal equilibration length obtained in the main text.

## S6. THEORETICAL MODEL OF THE DEVICE

### A. Setup

In this Section, we theoretically model an experimental device of type $B1$ and $B2$ in the main text, as depicted in Fig. S6. The device consists of three arms (labelled $1-3$), separated by insulating regions and connected to a central floating contact $\Omega_m$. Each arm is tuned to filling factor $\nu$ and the associated edge states have one incoming and one outgoing branch with respect to $\Omega_m$ and to the charge-flow direction (indicated by arrows in Fig. S6). We focus on such FQH states that exhibit counterpropagating edge modes, so that each branch hosts both downstream and upstream modes. Assuming efficient charge equilibration, non-equilibrium charge currents flow only downstream (in the direction of arrows), which is in our convention the counter-clockwise direction in each of the arms (see the chirality sign in the bottom right corner of the figure). By contrast, we assume that the thermal equilibration between edge channels is very weak. On arm 3, there are two electrodes U and D, with attached amplifiers $A_U$ and $A_D$ (where subscripts indicate the upstream and downstream locations with respect to $\Omega_m$), in which voltage/current fluctuations are measured. Except for $\Omega_m$, all electrodes are assumed to be at the base temperature $T_0$.

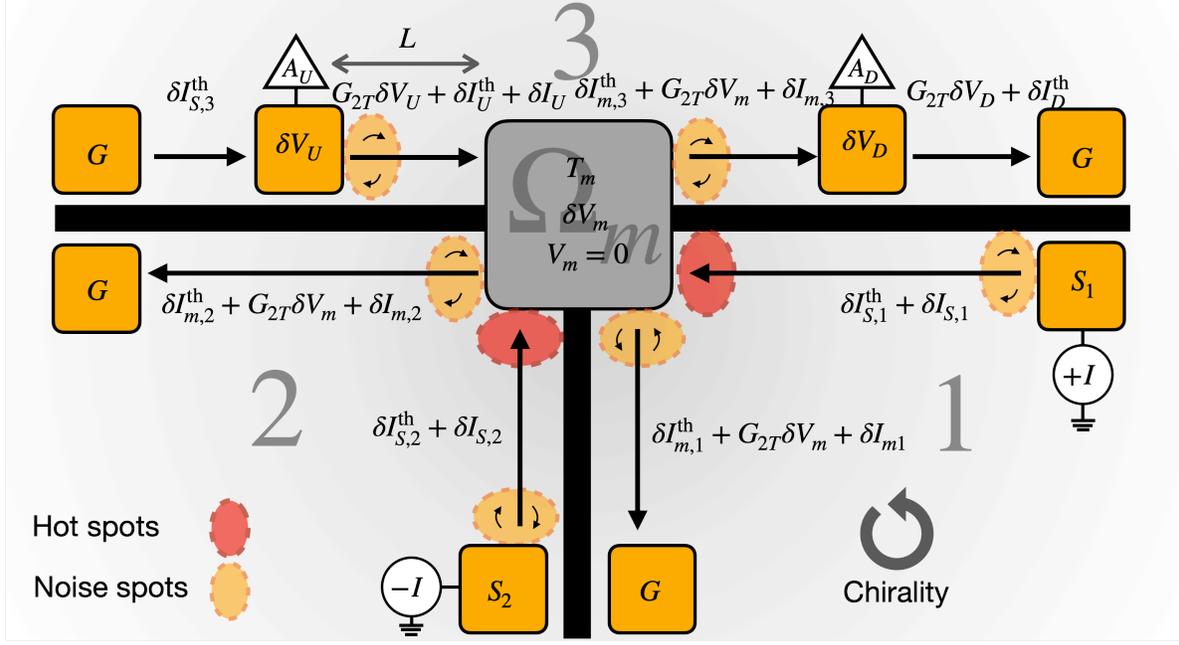

FIG. S6. Device schematics. The device consists of three arms (labels 1, 2, and 3), separated by insulating regions (black stripes) and connected to the central contact $\Omega_m$. Black arrows denote the direction of charge flow along the edge states in each arm. Impinging currents $\pm I$ heat up $\Omega_m$ by Joule heating from voltage drops at the hot spots (red regions) and inside $\Omega_m$. Partitioning of particle-hole pairs occurs at the noise spots (yellow regions). Excess noise is measured in contacts U (upstream from $\Omega_m$) and D (downstream from $\Omega_m$), with the help of amplifiers $A_U$ and $A_D$. The upstream propagation length, between $\Omega_m$ and $A_U$, is denoted $L$. Current and voltage fluctuations are labelled $\delta V$ and $\delta I$ respectively. For the meaning of specific fluctuations, see Secs. S6 B 1- S6 B 2.

The device is operated by the currents $I_1$ and $I_2$ biasing the arms 1 and 2 in source contacts $S_1$ and $S_2$, respectively. The source contact in the arm 3 is grounded, as marked by the label G in the figure. The two injected currents have equal magnitude but opposite sign: $I_1 = -I_2 \equiv I$. The average voltage $V_m$ of the central contact and the injected electrical power $P$ are then given as

$$V_m = \frac{I_1 + I_2}{2G_{2T}} = 0, \tag{S5}$$

$$P = 2 \times \frac{I^2}{2G_{2T}} = \frac{I^2}{G_{2T}}, \tag{S6}$$

where $G_{2T} = \nu e^2/h \equiv \nu G_0$. While $V_m = 0$, the floating electrode $\Omega_m$ is subject to voltage fluctuations $\delta V_m$.

In Fig. S6, we have also marked hot spots and noise spots. Hot spots are regions where the injected power dissipates and heat is generated. There are two such hot spots that heat $\Omega_m$; in addition, a part of the heat is dissipated inside $\Omega_m$. The noise spots are regions close to contacts where partitioning of charge due to edge impurity scattering may give rise to excess dc noise[S10,S11]. This happens if heat from hot spots or heated contacts reaches the noise spot. Due to the chiral nature of the edge and efficient charge equilibration, partitioning by scattering in regions other than noise spots does not contribute to the excess dc noise.

Details of the theoretical analysis of heat transport and noise in this setup are presented in the subsequent Sections. More specifically:

a) In Sec. S7 we develop a microscopic model for the computation of noise, under the assumption of vanishing thermal equilibration between edge channels, and derive formulas for the noise on an edge segment connecting two contacts with different temperatures.

b) In Sec. S8 we use the results of Sec. S7 to compute the downstream noise, i.e., the excess noise $S_{\text{excess}}^{\text{D}}$ in the downstream contact D (with amplifier $A_{\text{D}}$). This result allows us extract the central-contact temperature $T_m$ from experimental measurements of the noise.

c) In Sec. S9 we use the results of Sec. S7 to compute the upstream noise, i.e., the excess noise $S_{\text{excess}}^{\text{U}}$ in the upstream electrode U (with amplifier $A_{\text{U}}$) as a function of $T_m$.

d) In Sec. S10 we establish a power-balance relation between the injected electrical power $P$ and the outgoing edge heat current $J_{\text{edge}}^{Q} = 3\kappa_{2T}(T_m^2 - T_0^2)/2$. This relation, in combination with the results for the noise, allows us to extract experimentally the heat conductance $G_{2T}^{Q} = \kappa_{2T} T$ of the device (which is naturally measured in units of $\kappa_0 = \pi^2 k_B^2/3h$).

e) In Sec. S11 we discuss theoretical predictions for the heat conductance $\kappa_{2T}/\kappa_0$ in the regime of vanishing heat equlibration.

f) A comparison of results from the theoretical analysis with the experiment is presented in Sec. S12.

## B. General analysis of downstream and upstream noise

As a starting point for the theory, we derive here expressions for the noise in the two amplifiers from a general analysis of the device.

### 1. Downstream noise

We begin by computing the downstream noise. We note first that, since we are interested in the noise at a low frequency $\omega_m$, the condition of charge conservation at the central Ohmic contact $\Omega_m$ should be imposed. The charge conservation holds under the condition $(h/e^2)\omega_m C \ll 1$, where $C$ is the capacitance of the island. Equivalently, this condition can be written as $\omega_m \ll E_C$, where $E_C \sim e^2/C$ is the charging energy of $\Omega_m$. Let us emphasize that, at the same time, the capacitance $C$ is assumed to be sufficiently large such that $E_C \ll k_B T_0$. This ensures that $\Omega_m$ efficiently equilibrates impinging edge channels, so that heat Coulomb blockade[S3,S4] is not operative.

The incoming and outgoing current fluctuations on $\Omega_m$ are (see Fig. S6)

$$\delta I_m^{\text{in}} = \left(\delta I_{S_1}^{\text{th}} + \delta I_{S_1}\right) + \left(\delta I_{S_2}^{\text{th}} + \delta I_{S_2}\right) + \delta I_U^{\text{th}} + \delta I_U - \delta I_{m,1} - \delta I_{m,2} - \delta I_{m,3}, \quad (S7)$$

$$\delta I_m^{\text{out}} = 3G_{2T}\delta V_m + \delta I_{m,1}^{\text{th}} + \delta I_{m,2}^{\text{th}} + \delta I_{m,3}^{\text{th}}. \quad (S8)$$

In Eqs. (S7) and (S8), $\delta I_{S_i}$ are non-equilibrium current fluctuations impinging on $\Omega_m$. These contributions originate from heat back-propagating from $\Omega_m$ to the noise spots (yellow regions) close to the source contacts[S10,S11]. Likewise, $\delta I_{m,i}$ and $\delta I_U$ come from charge partitioning in the noise spots close to the central contact $\Omega_m$ and the upstream contact U. Finally, $\delta I_{S_i}^{\text{th}}$, $\delta I_{m,i}^{\text{th}}$, and $\delta I_U^{\text{th}}$ are equilibrium current fluctuations from sources, the central contact $\Omega_m$, and the upstream contact U, respectively. Their noise correlations are given by Johnson-Nyquist noise with their respective temperature.

As discussed in Sec. S1, the upstream contact is assumed to be grounded at the frequency range where downstream noise is measured. Then, the current fluctuations from the upstream contact U read $\delta I_U^{\text{th}} + \delta I_U$, which appear in Eq. (S7).

Equating (in view of charge conservation) $\delta I_m^{\text{in}} = \delta I_m^{\text{out}}$ and solving for $G_{2T}\delta V_m$, we find

$$G_{2T}\delta V_m = \frac{1}{3}\left(\Delta I_{S_1} + \Delta I_{S_2} + \Delta I_U - \Delta I_{m,1} - \Delta I_{m,2} - \Delta I_{m,3}\right), \quad (S9)$$

where $\Delta I_i \equiv \delta I_i^{\text{th}} + \delta I_i$ for $i = (S_1), (S_2), (m,1), (m,2), (m,3)$, or $U$. Next, by equating incoming and outgoing fluctuations at the downstream contact D, we find

$$\left(\delta I_{m,3}^{\text{th}} + \delta I_{m,3}\right) + G_{2T}\delta V_m = G_{2T}\delta V_D + \delta I_D^{\text{th}}, \tag{S10}$$

where $\delta V_D$ are the local voltage fluctuations and $I_D^{\text{th}}$ are thermal fluctuations associated with the temperature $T_0$ of the contact D.

The excess downstream noise is defined as

$$S_{\text{excess}}^{\text{D}} = \overline{(G_{2T}\delta V_D)^2} - 4G_{2T}k_B T_0. \tag{S11}$$

By inserting $G_{2T}\delta V_m$ from Eq. (S9) into Eq. (S10), solving for $G_{2T}\delta V_D$, and substituting in Eq. (S11), we obtain

$$\begin{aligned}
S_{\text{excess}}^{\text{D}} &= \frac{4}{9}\overline{(\Delta I_{m,3})^2} + \overline{\left(\delta I_D^{\text{th}}\right)^2} + \frac{1}{9}\left[\overline{(\Delta I_{m,1})^2} + \overline{(\Delta I_{m,2})^2} + \overline{(\Delta I_{S_1})^2} + \overline{(\Delta I_{S_2})^2} + \overline{(\Delta I_U)^2}\right] \\
&\quad - 4G_{2T}k_B T_0 \\
&= \frac{2}{3}\overline{(\Delta I_m)^2} + \frac{1}{9}\left[S_{\text{excess}}^{S_1} + S_{\text{excess}}^{S_2} + S_{\text{excess}}^{\text{U}}\right] - \frac{4}{3}G_{2T}k_B T_0.
\end{aligned} \tag{S12}$$

Here we have used independence of different sources of fluctuations, which implies that the cross-correlations are zero. In the final equality, we assumed that $\overline{(\Delta I_{m,j})^2} \equiv \overline{(\Delta I_m)^2}$ are independent of $j$, i.e., equal for all three arms. We have also defined excess noises $S_{\text{excess}}^i \equiv \overline{(\Delta I_i)^2} - 2G_{2T}k_B T_0$ for $i = (S_1), (S_2)$, and $U$. Furthermore, $\overline{\left(\delta I_D^{\text{th}}\right)^2} = 2G_{2T}k_B T_0$.

Let us analyze the final form of Eq. (S12). The first two terms there represent two distinct contributions to the downstream noise. The first term is the thermal noise of $\Omega_m$ (see Sec. S8 for a detailed discussion), while the second term results from the sum of excess noises from partitioning close to the upstream amplifier, $S_{\text{excess}}^{\text{U}}$ (see Secs. S6 B 2 and S9 for detailed discussions) and the source contacts, $S_{\text{excess}}^{S_j}$. The last term (with the minus sign) is the subtraction of thermal noise (with temperature $T_0$), in correspondence with the definition of the excess noise.

If edge channels are thermally equilibrated (e.g., by impurity scattering), the Johnson-Nyquist relation holds for $\overline{(\Delta I_m)^2}$, yielding $\overline{(\Delta I_m)^2} = 2G_{2T}k_B T_m$. In this case we arrive at

$$S_{\text{excess}}^{\text{D}} = \frac{4}{3}k_B G_{2T}(T_m - T_0) + \frac{1}{9}\left[S_{\text{excess}}^{S_1} + S_{\text{excess}}^{S_2} + S_{\text{excess}}^{\text{U}}\right]. \tag{S13}$$

Equation (S13) establishes a relation between the downstream noise and $\Delta T \equiv T_m - T_0$. Therefore, it allows one to determine the temperature of the central contact $\Omega_m$ by measuring the downstream noise $S^D_{\text{excess}}$. The last term in Eq. (S13) is a relatively small correction (due to the factor 1/9). Still, it is appreciable and should be taken into account if one wants to find $T_m$ with a good accuracy. This is done in the present work, when the formula generalizing Eq. (S13) on a non-equilibrated regime is used for determining $T_m$. The noise $S^U_{\text{excess}}$ is directly measured experimentally. The noises $S^{S_1}_{\text{excess}}$ and $S^{S_2}_{\text{excess}}$ can be also be obtained from measurements of $S^U_{\text{excess}}$, since the generation of upstream noise is fully analogous in all three arms.

In Sec. S8, we microscopically compute $\overline{(\Delta I_m)^2}$ in Eq. (S12) for the case of the absence of thermal equilibration at the edge. We find how the Johnson-Nyquist formula for $\overline{(\Delta I_m)^2}$ should be corrected when the edge is not thermally equilibrated. This allows us to extend on such non-equilibrated regime the procedure of determining the temperature $T_m$ by means of measuring the downstream noise $S^D_{\text{excess}}$.

### 2. Upstream noise

We now consider the upstream noise on arm 3, as measured in the contact U by means of the amplifier $A_U$. Conservation of current fluctuations at the upstream contact U gives

$$-\delta I^U + \delta I^{\text{th}}_{S_3} = G_{2T}\delta V_U + \delta I^{\text{th}}_U, \tag{S14}$$

where $\delta I^{\text{th}}_{S_3}$ are equilibrium fluctuations impinging from the top left grounded contact. Here we have taken into account that, under the used experimental design, the upstream contact U is floating at the frequency range where upstream noise is measured. Re-arranging, we get $G_{2T}\delta V_U = \delta I^{\text{th}}_{S_3} - \delta I^{\text{th}}_U - \delta I^U = \delta I^{\text{th}}_{S_3} - \Delta I_U$, so that the upstream excess noise is given by

$$S^U_{\text{excess}} \equiv \overline{(G_{2T}\delta V_U)^2} - 4k_B G_{2T} T_0 = \overline{(\Delta I_U)^2} + \overline{(\delta I^{\text{th}}_{S_3})^2} - 4G_{2T}k_B T_0$$
$$= \overline{(\Delta I_U)^2} - 2G_{2T}k_B T_0. \tag{S15}$$

Here, we used that all cross-correlations are zero and $\overline{\left(\delta I^{\text{th}}_{S_3}\right)^2} = 2G_{2T}k_B T_0$. In Eq. (S15), $\overline{(\Delta I_U)^2}$ is the non-equilibrium noise generated by partitioning at the noise spot just to the right of the contact U. Equation (S15) states that the upstream amplifier detects noise which is generated when heat flows from $\Omega_m$ and heats the noise spot. Hence, the upstream noise can be used as a local thermometer of the noise spot. Under the assumption that the contact U is acting as a heat reservoir (i.e., its temperature is kept fixed at $T_0$), the upstream excess noise is independent of other non-equilibrium noise sources in the device. In Sec. S9, we compute $\overline{(\Delta I_U)^2}$ microscopically.

## S7. MICROSCOPIC MODEL AND CALCULATION OF NOISE ON AN EDGE SEGMENT

In the preceding Section, we derived general formulas for the excess noise. The key quantities there are $\overline{(\Delta I_m)^2}$ in Eq. (S12) for the downstream noise and $\overline{(\Delta I_U)^2}$ in Eq. (S15) for the upstream noise. Both these quantities represent a noise generated on an edge segment connecting two contacts with different temperatures: $T_0$ and $T_m$, with $T_m > T_0$. The only difference is that in the case of $\overline{(\Delta I_m)^2}$ the upstream contact $\Omega_m$ is hot and the downstream contact D is cold, while in the case of $\overline{(\Delta I_U)^2}$ the situation is opposite: the upstream contact U is cold, while the downstream contact $\Omega_m$ is hot. We are now going to compute the noise generated on a segment between the contacts with two different temperatures, $T_L$ and $T_R$. For $T_L > T_R$ this will give $\overline{(\Delta I_m)^2}$, and for $T_L < T_R$ we will find $\overline{(\Delta I_U)^2}$.

To this end, we present in this Section a microscopic model to compute noise generated on a FQH edge segment connected to two contacts at $x = 0$ and $x = L$ (see Fig. S7a). To do so, we use a formalism developed in Refs. S10–S13. In accordance with experimental observations, we always assume full charge equilibration, $l_{eq}^C \ll L$, where $l_{eq}^C$ is the characteristic length scale (charge equilibration length) of inter-edge channel electron tunneling. However, in contrast to previous work[S14], we are here interested in noise in the regime of no thermal equilibration between the edge channels. Weak thermal equlibration can be theoretically achieved with sufficiently strong inter-channel interactions[S15]. We first focus on the $\nu = 2/3$ state, which is the archetypical hole-conjugate state, and whose edge hosts channels of both chiralities. Our approach is then generalized to other hole-conjugate states.

### A. Filling 2/3

The edge segment consists of two counter-propagating edge channels, one downstream (+) associated with the corresponding filling factor discontinuity $\nu_+$ and one upstream channel (-) associated with $\nu_-$. In contrast to the standard Luttinger liquid ($\nu_+ = \nu_- = 1$), the chiral nature of the edge is captured by $\nu_+ > \nu_-$; downstream transport is taken from left to right. Specifically, $\nu_+ = 1$ and $\nu_- = 1/3$ for the $\nu = 2/3$ edge[S16]. We divide the the edge segment into three regions: the left contact region, a central region, and the right contact region (see Fig. S7b). While the inter-channel interaction is assumed to be screened to zero in the contacts, it is finite in the central region (see Ref. S17 for a detailed discussion on this assumption). The scaling dimension $\Delta$ of the inter-channel tunneling operator (to be discussed below) quantifies the strength of the inter-channel interaction; $\Delta = 2$ corresponds to non-interacting channels while $\Delta = 1$ corresponds to the regime of strongly interacting channels[S16]. Fig. S7b shows $\Delta(x)$ as a function of position $x$ due to the attached contacts. We assume that $\Delta$ varies sharply at the boundaries between the regions, which is justified at low frequencies[S6]. The two contacts are taken at different temperatures $T_L$ and $T_R$, but at the same electrochemical potential.

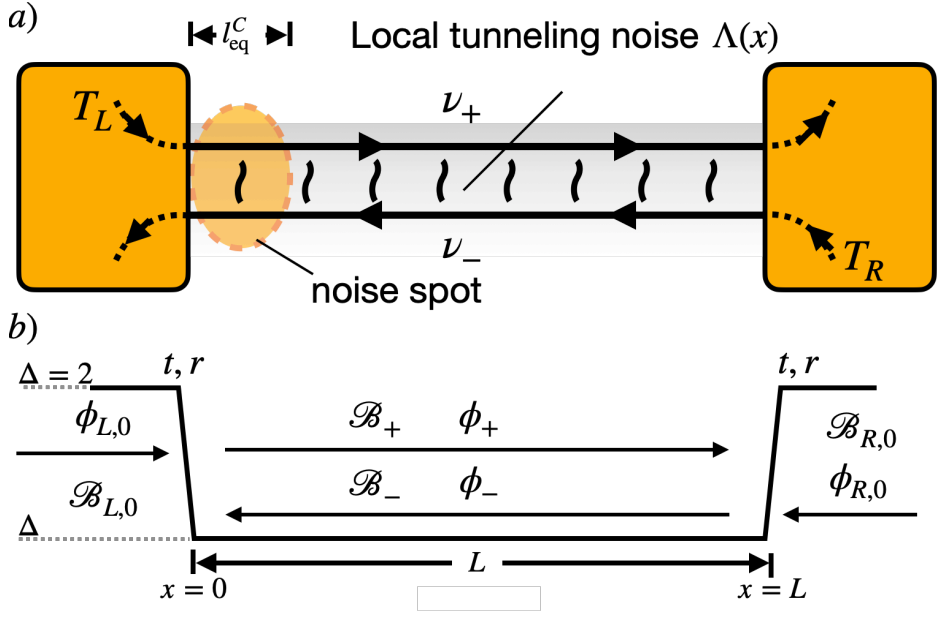

FIG. S7. Noise generation on the edge. a) The two contacts are at different temperatures $T_L$ and $T_R$ but at the same voltage. Excess dc noise is predominantly generated in the region of the spatial extent $\sim l_{\rm eq}^C$ (noise spot) close to the left contact, see Eq. (S16). b) Bosonic scattering states $\phi_{L,0}$ and $\phi_{R,0}$ emanating from the left and right contacts are characterized by equilibrium distribution functions $\mathcal{B}_{L,0}$ and $\mathcal{B}_{R,0}$ controlled by temperatures of the contacts. The distribution functions $\mathcal{B}_\pm$ of the eigenmode states $\phi_\pm$ in the interacting region are determined by $\mathcal{B}_{L,0}$ and $\mathcal{B}_{R,0}$ in combination with reflection ($r$) and transmission ($t$) amplitudes of bosons at boundaries between the contacts and the interacting region, see Eqs. (S30)-(S31).

The zero frequency noise $S$ in either of the two contacts is generally given as[S11]

$$S = \frac{2e^2}{hl_{\text{eq}}^C} \frac{\nu_-}{\nu_+}(\nu_+ - \nu_-) \int_0^L dx\, \Lambda(x) e^{-\frac{2x}{l_{\text{eq}}^C}} + \frac{2e^2}{h} k_B T_L \frac{(\nu_+ - \nu_-)^2}{\nu_+}. \tag{S16}$$

Here, the first term describes the noise resulting from inter-channel electron tunneling along the edge. The exponential factor in the integral is a result of the chiral charge transport, and implies that the dominant noise contribution comes from a region of size $\sim l_{\text{eq}}^C$ close to the left contact. We call this region the noise spot. The local noise generated by impurity scattering is described by the noise kernel $\Lambda(x)$ given by

$$\Lambda(x) \equiv \frac{S_{\text{loc}}(x, T_L, T_R)}{2g_{\text{loc}}(x, T_L, T_R)}, \tag{S17}$$

where $S_{\text{loc}}(x, T_L, T_R)$ and $g_{\text{loc}}(x, T_L, T_R)$ is the local electron tunneling dc noise and the tunneling conductance, respectively. Most generically, $S_{\text{loc}}$ and $g_{\text{loc}}$ depend on microscopic details such as the inter-channel interaction, the edge disorder strength, the local voltage difference between the channels, and the edge channel energy distributions.

The second term in Eq. (S16) describes downstream propagating thermal fluctuations injected by the left contact. The thermal fluctuations injected by the right contact reaching the noise spot are suppressed by a factor that is exponential in $L/l_{\text{eq}}^C \gg 1$ and thus can be safely neglected[S10].

In the present work, we are interested in the case of no voltage bias and no thermal equilibration. In this situation, Eq. (S16) can be simplified. Specifically, $\Lambda$ becomes independent of $x$, and the integration can be trivially performed. Further, $\Lambda$ depends only on the temperatures $T_L$ and $T_R$ and the interaction parameter $\Delta$. To emphasize this, we use the notation $\Lambda(T_L, T_R, \Delta)$ for the noise kernel. Thus, Eq. (S16) reduces to

$$S = \frac{e^2}{h}\frac{\nu_-}{\nu_+}(\nu_+ - \nu_-)\Lambda(T_L, T_R, \Delta) + \frac{2e^2}{h} k_B T_L \frac{(\nu_+ - \nu_-)^2}{\nu_+}. \tag{S18}$$

Our goal is now to use Eqs. (S17) and (S18) to compute the excess charge noise generated by the temperature difference, $T_L \neq T_R$, in the absence of a net charge current flow. This type of noise is called "$\Delta T$ noise", and has in recent years attracted increasing attention, both from theoretical[S18–S24] and experimental[S25–S28] points of view.

Equations (S17) and (S18) are applicable to both downstream and upstream noise (see Fig. S6). As discussed above, the upstream noise corresponds to $T_L = T_0$ and $T_R = T_m$, i.e., to the right contact being hot, $T_R > T_L$ (the edge segment on arm 3 to the left of $\Omega_m$ in Fig. S6). Conversely, the downstream noise corresponds to $T_L = T_m$ and $T_R = T_0$, i.e., to a hot left contact, $T_L > T_R$ (the edge segment on arm 3 to the right of $\Omega_m$ in Fig. S6).

The expectation values of the local noise and electron tunneling conductance can be computed within the chiral Luttinger liquid model[S29]. To leading order in the tunneling strength $\Gamma_0$ (that is assumed for simplicity uniform), these quantities are given as[S15,S21]

$$S_{\text{loc}} = 4 \int_{-\infty}^{\infty} d\tau \langle \mathcal{T}(\tau,0)\mathcal{T}^\dagger(0,0)\rangle, \tag{S19}$$

$$g_{\text{loc}} = 2i \int_{-\infty}^{\infty} d\tau \tau \langle \mathcal{T}(\tau,0)\mathcal{T}^\dagger(0,0)\rangle. \tag{S20}$$

Here

$$\mathcal{T}(\tau,0) = \frac{\Gamma_0}{2\pi b} \exp\left[i\sqrt{\Delta-1}\phi_+(\tau,0) + i\sqrt{\Delta+1}\phi_-(\tau,0)\right] \tag{S21}$$

is the local tunneling operator expressed in terms of the bosonic eigenmodes $\phi_\pm$ in the interacting region[S15], and $b$ is a short distance cutoff. Evaluating the correlation function, we find

$$\langle \mathcal{T}(\tau,0)\mathcal{T}^\dagger(0,0)\rangle = \frac{|\Gamma_0|^2}{(2\pi b)^2} G_+(\tau,0)^{\Delta-1} G_-(\tau,0)^{\Delta+1}, \tag{S22}$$

where

$$G_\pm(\tau,0) = \exp\left[\langle \phi_\pm(\tau,0)\phi_\pm(0,0)\rangle - \langle \phi_\pm(0,0)\phi_\pm(0,0)\rangle\right] \equiv \exp\left[\mathcal{B}_\pm(\tau,0)\right]. \tag{S23}$$

The Fourier transform of $\mathcal{B}_\pm(\tau,0)$ in frequency space, $\mathcal{B}_\pm(\omega,0)$, corresponds to the distribution functions of the eigenmodes $\phi_\pm$. In the absence of thermal equilibration, these are non-equilibrium distribution functions, arising from scattering of bosonic modes at the boundaries between contacts and the interacting middle region (see Fig. S7b). We therefore now seek to express $\mathcal{B}_\pm$ in terms of the known equilibrium Bose distributions $\mathcal{B}_{L/R,0}$ of the non-interacting bosons $\phi_{L/R,0}$ in the contacts. Our approach follows closely that in Refs. S6 and S30. The bosonic states in the central region $\phi_\pm(\tau,x)$ can be expanded as

$$\phi_+(\tau,x) = t \sum_n r^{2n} \phi_{L,0}(\tau - 2n\tilde{\tau}, x) + tr \sum_n r^{2n} \phi_{R,0}(\tau - 2n\tilde{\tau}, x),$$

$$\phi_-(\tau,x) = t \sum_n r^{2n} \phi_{R,0}(\tau - 2n\tilde{\tau}, x) + tr \sum_n r^{2n} \phi_{L,0}(\tau - 2n\tilde{\tau}, x). \tag{S24}$$

Here, the parameter $\tilde{\tau} = \frac{L}{2}(v_+^{-1} + v_-^{-1})$ is the mean "flight-time" through the interacting region[S6] and $v_+$ and $v_-$ are the velocities of $\phi_\pm$. The parameters $t$ and $r$ are transmission and reflection coefficients due to the sharp change in interaction strength at the boundaries. The coefficients are given as

$$t = \frac{2(1-c^2)^{1/4}}{\sqrt{1-c} + \sqrt{1+c}}, \tag{S25}$$

$$r = \frac{1 - \sqrt{1-c^2}}{c}, \tag{S26}$$

where $c$ is related to $\Delta$ as

$$\Delta = \frac{2 - \sqrt{3}c}{\sqrt{1-c^2}}. \tag{S27}$$

It follows from Eq. (S24) that

$$\mathcal{B}_\pm(\omega,0) = t^2 \sum_{n,m}^\infty r^{2(n+m)} e^{2i(m-n)\omega\tilde{\tau}} \mathcal{B}_{L/R}(\omega,0) + t^2 r^2 \sum_{n,m}^\infty r^{2(n+m)} e^{2i(m-n)\omega\tilde{\tau}} \mathcal{B}_{R/L}(\omega,0), \tag{S28}$$

where $\mathcal{B}_{R/L}(\omega,0)$ is the Fourier transform of

$$\mathcal{B}_{R/L}(\tau,0) = \langle \phi_{R/L}(\tau,0)\phi_{R/L}(0,0)\rangle - \langle \phi_{R/L}(0,0)\phi_{R/L}(0,0)\rangle. \tag{S29}$$

We now neglect terms of the form $e^{2im\tilde{\tau}}$ which arise due to the Fabry-Perot interference of bosonic modes reflected at boundary. These terms lead to an oscillatory behavior in energy on the scale $\pi v_\pm/L$. These oscillations will however be suppressed at temperatures $T_L, T_R \gg \pi v_\pm/L$, which is assumed as follows. In this case, Eq. (S28) reduces to

$$\mathcal{B}_+(\omega,0) = \frac{T}{1-R^2}\mathcal{B}_L(\omega,0) + \frac{TR}{1-R^2}\mathcal{B}_R(\omega,0), \tag{S30}$$

$$\mathcal{B}_-(\omega,0) = \frac{T}{1-R^2}\mathcal{B}_R(\omega,0) + \frac{TR}{1-R^2}\mathcal{B}_L(\omega,0). \tag{S31}$$

with $T = t^2$, $R = r^2$. Substituting the relations (S30)-(S31) into Eq. (S23), and the result into Eq. (S22), we express the tunneling correlation function as

$$\langle \mathcal{T}(\tau,0)\mathcal{T}^\dagger(0,0)\rangle = \frac{|\Gamma_0|^2}{(2\pi b)^2} G_L(\tau,0)^{2d_L} G_R(\tau,0)^{2d_R}, \tag{S32}$$

where $G_L(\tau,0)$ and $G_R(\tau,0)$ are the equilibrium free-fermion Green's functions,

$$G_{L,R}(\tau,0) = \frac{\pi b T_{L,R}/v_{L,R}}{\sin\left[\frac{\pi T_{L,R}}{v_{L,R}}(b - i\tau v_{L,R})\right]}. \tag{S33}$$

Here, $v_{L,R}$ are the velocities of the bosonic modes in the contacts. At zero temperature, $G_{L,R}$ reduce to

$$G_{L,R}(\tau, 0) = \frac{b}{b - i\tau v_{L,R}}. \tag{S34}$$

The exponents $d_L$ and $d_R$ in Eq. (S32) capture the interaction dependence of the tunneling. They are given in terms of $\Delta$ by the following expressions:

$$2d_L = \frac{\Delta^3 + \Delta - \sqrt{3(\Delta^2 - 1)}}{\Delta^2 + 3}, \tag{S35}$$

$$2d_R = \frac{\Delta^3 + 5\Delta + \sqrt{3(\Delta^2 - 1)}}{\Delta^2 + 3}. \tag{S36}$$

The Fourier transforms of the factors $G_{L,R}(\tau, 0)^{2d_{L,R}}$ read

$$P_{L,R}(\omega, T_{L,R}) = \int_{-\infty}^{\infty} d\tau e^{i\omega\tau} G_{L,R}(\tau, 0)^{2d_{L,R}}$$
$$= \left(\frac{2\pi b T_{L,R}}{v_{L,R}}\right)^{2d_{L,R}-1} \left(\frac{b}{v_{L,R}}\right) e^{\omega/(2T_{L,R})} \frac{|\Gamma(d_{L,R} + i\frac{\omega}{2\pi T_{L,R}})|^2}{\Gamma(2d_{L,R})}. \tag{S37}$$

In the zero-temperature limit, $T_{L,R} \to 0$, they reduce to

$$P_{L,R}(\omega, 0) = \int_{-\infty}^{\infty} d\tau e^{i\omega\tau} G_{L,R}(\tau, 0)^{2d_{L,R}} = \frac{2\pi (b/v_{L,R})^{2d_{L,R}} \omega^{2d_{L,R}-1} \Theta(\omega)}{\Gamma(2d_{L,R})}. \tag{S38}$$

Here, $\Gamma(z)$ is the gamma-function, and $\Theta(\omega)$ is the Heaviside step function.

With the Fourier transforms, the noise (S19) can be expressed as

$$S_{\text{loc}} = \frac{4|\Gamma_0|^2}{(2\pi b)^2} \int_{-\infty}^{\infty} \frac{d\omega}{2\pi} P_L(-\omega, T_L) P_R(\omega, T_R), \tag{S39}$$

and the tunneling conductance (S20) as

$$g_{\text{loc}} = \frac{2|\Gamma_0|^2}{(2\pi b)^2} \frac{\partial}{\partial \omega'} \left( \int_{-\infty}^{\infty} \frac{d\omega}{2\pi} P_L(\omega' - \omega, T_L) P_R(\omega, T_R) \right) \bigg|_{\omega'=0}. \tag{S40}$$

Combining Eqs. (S39) and (S40), we express the noise kernel (S17) as

$$\Lambda(T_L, T_R, \Delta) = \frac{\int_{-\infty}^{\infty} d\omega P_L(-\omega, T_L) P_R(\omega, T_R)}{\frac{\partial}{\partial \omega'} \left( \int_{-\infty}^{\infty} d\omega P_L(\omega' - \omega, T_L) P_R(\omega, T_R) \right) \bigg|_{\omega'=0}}. \tag{S41}$$

Let us briefly recapitulate the procedure leading to Eq. (S41) for the noise kernel $\Lambda$. In this equation, the numerator and the denominator are, respectively, the local noise and the conductance of electron tunneling between two edge channels out of equilibrium. The bosonic scattering theory has allowed us to express these quantities in terms of known equilibrium distributions in the contacts. We recall that the situation of no thermal equilibration considered here is complementary to the limit of strong thermal equilibration, for which the local noise kernel can be expressed in terms of local edge channel temperatures[S10,S12]. Let us further emphasize that $\Lambda$ depends on the edge interaction parameter $\Delta$ through the exponents $2d_{L,R}$ [see Eqs. (S35)-(S36)] in the expressions for $P_{L,R}$ [see Eq. (S37)-(S38)].

Substituting the expression (S41) into the noise formula (S18), we arrive at the final expression for the noise on the 2/3 edge:

$$S = \frac{2}{9}\frac{e^2}{h} \times \Lambda(T_L, T_R, \Delta) + \frac{8}{9}\frac{e^2}{h}k_B T_L. \quad (S42)$$

It is instructive to inspect Eq. (S42) in the case of global equilibrium, $T_L = T_R = T_0$. In this situation, we can analytically perform the integrals in Eq. (S41) to obtain $\Lambda(T_0, T_0, \Delta) = 2k_B T_0$, independent of $\Delta$. Then, Eq. (S42) reduces (as expected) to the Nyquist-Johnson relation

$$S = 2G_{2T}k_B T_0, \quad G_{2T} = \frac{2}{3}\frac{e^2}{h}. \quad (S43)$$

Further, we briefly consider the regime of strong thermal equilibration. Then both edge channels equilibrate to $T_L$ at the noise spot. In this case, the measured downstream noise is the Johnson-Nyquist noise at temperature $T_L$, i.e,

$$S = 2G_{2T}k_B T_L, \quad G_{2T} = \frac{2}{3}\frac{e^2}{h}. \quad (S44)$$

### B. Filling 3/5

The edge at filling $\nu = 3/5$ consists of three modes. One downstream mode $\phi_1$ with filling factor discontinuity $\nu_1 = 1$ and two upstream modes $\phi_{1/3}$, $\phi_{1/15}$ with $\nu_{1/3} = 1/3$ and $\nu_{1/15} = 1/15$ respectively. To greatly simplify the noise analysis of this edge, we now make the following assumptions. First, we assume that the equilibration lengths (for both charge and heat) of the two upstream modes is very small. We thus "merge" these two modes into one effective upstream mode. Secondly, we assume that both inter-mode tunneling and the inter-mode interaction is dominated by that between $\phi_1$ and $\phi_{1/3}$. This assumption is based on that these modes are spatially closer than $\phi_1$ and $\phi_{1/15}$. Additionally, tunneling and interactions between the co-propagating $\phi_{1/3}$ and $\phi_{1/15}$ does not influence the equilibration or the heat conductance. With these simplifications, the noise kernel in Eq. (S16) becomes the same for $\nu = 2/3$ and $\nu = 3/5$ since they are based on the same single tunneling operator [see Eq. (S21)]. The difference in the generated noise for these fillings, in the thermally non-equilibrated regime, is the values of $\nu_\pm$. While $\nu = 2/3$, we have $\nu_+ = 1$ and $\nu_- = 1/3$, for $\nu = 3/5$, we have instead $\nu_+ = 1$ and $\nu_- = 1/3 + 1/15 = 2/5$. With these parameters, we have obtain the noise for $\nu = 3/5$ as

$$S = \frac{6}{25}\frac{e^2}{h} \times \Lambda(T_L, T_R, \Delta) + \frac{18}{25}\frac{e^2}{h}k_B T_L. \tag{S45}$$

For global equilibrium, we recover $S = 2G_{2T}k_B T_0$ with $G_{2T} = (3/5)e^2/h$, and for full thermal equilibration at temperature $T_L$ in the noise spot, we find $S = 2G_{2T}k_B T_L$.

In the following two Sections, we use Eqs. (S42) and (S45) to compute downstream and upstream noise in the case of no thermal equilibration.

## S8. CALCULATION OF DOWNSTREAM NOISE

In this Section, we use the results of Sec. S7 to compute the downstream noise on an edge segment. The downstream noise configuration corresponds to choosing $T_L = T_m = T_0 + \Delta T$ and $T_R = T_0$ with $\Delta T > 0$. We may then identify the noise in Eq. (S42) with the downstream noise $\overline{(\Delta I_m)^2}$ from Eq. (S12). We then have for $\nu = 2/3$

$$\overline{(\Delta I_m)^2} = \frac{2}{9}\frac{e^2}{h} \times \Lambda(\Delta T + T_0, T_0, \Delta) + \frac{8}{9}\frac{e^2}{h}k_B(\Delta T + T_0). \tag{S46}$$

In Fig. S8a, we plot $\overline{(\Delta I_m)^2}$ as a function of $\Delta T$ for $T_0 = 14$ and various values of $\Delta$. We choose this value of the base temperature $T_0$ for convenience of comparison to our experiment: $T_0$ was measured to be 14 mK for $\nu = 2/3$. We set the prefactor $(e^2/h)k_B$ in the expression for $\overline{(\Delta I_m)^2}$ to be unity. As a result, $\overline{(\Delta I_m)^2}$ is measured in the same units as the temperature.

We first note that the $\Delta$-dependence of $\Lambda(T_0 + \Delta T, T_0, \Delta)$ is rather weak, within a few percent. It becomes especially weak for not too large $\Delta T$, i.e., $\Delta T \lesssim 30$. In this regime, an excellent approximation to $\Lambda(T_0 + \Delta T, T_0, \Delta)$ is obtained by expanding Eq. (S42) to first order in $\Delta T$ for $\Delta = 1$. The result is the simple expression

$$\Lambda(T_0 + \Delta T, T_0, \Delta) \approx 2T_0 + 0.5\Delta T. \tag{S47}$$

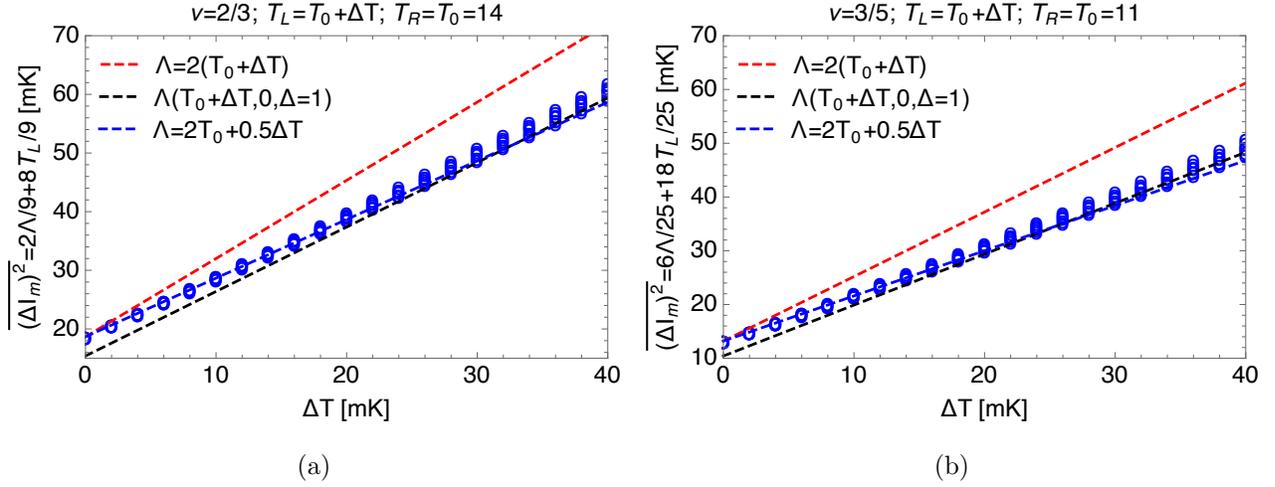

FIG. S8. Downstream noise, $\overline{(\Delta I_m)^2}$ [see Eqs. (S46) and (S48)], as functions of $\Delta T = T_L - T_0$ in units of mK. We set $e^2/h = k_B = 1$ so that $\overline{(\Delta I_m)^2}$ is also measured in mK. (a) For $\nu = 2/3$, with $T_0 = 14$ corresponding to the experimentally measured value $T_0 = 14$ mK at this filling. The blue circles correspond to $\Delta$ varying from 1 (bottom) to 2 (top). The blue dashed line, Eq. (S47), is an excellent approximation for $0 \lesssim \Delta T \lesssim 30$. The black dashed line is obtained by using the large-$\Delta T$ asymptotics $\Lambda(\Delta T + T_0, 0, \Delta = 1) \simeq 0.948(\Delta T + T_0)$. The red dashed line is Johnson-Nyquist noise with the temperature $T_0 + \Delta T$ [see Eq. (S44)], for which $\overline{(\Delta I_m)^2} = 4(\Delta T + T_0)/3 = 2G_{2T}(T_0 + \Delta T)$ with $G_{2T} = 2/3$; this corresponds to the limit of full thermal equilibration. (b) Similar to (a) but at $\nu = 3/5$, with $T_0 = 11$ corresponding to the measured $T_0 = 11$ mK at this filling. The red, dashed line is Johnson-Nyquist noise $2G_{2T}(T_0 + \Delta T)$ with $G_{2T} = 3/5$.

The corresponding result for the noise $\overline{(\Delta I_m)^2}$ is shown by the blue dashed line in Fig. S8a. In the same plot, we present (by black dashed line) the result for $\overline{(\Delta I_m)^2}$ obtained by using the large-$\Delta T$ asymptotics $\Lambda(\Delta T + T_0, 0, \Delta = 1) \simeq 0.948(\Delta T + T_0)$, which is a very good approximation for $\Delta T \gtrsim 20$. Further, the red line displays the Nyquist-Johnson noise $2G_{2T}k_B(\Delta T + T_0)$, which is obtained for $\overline{(\Delta I_m)^2}$ in the limit of full thermal equilibration, see Eq. (S44). Quite remarkably, the difference between the limits of no equilibration and full equilibration is rather small, within $\approx 20\%$. In the intermediate case of partial thermal equilibration, the results are expected to be in between. This weak dependence of the downstream noise on the degree of thermal equilibration, in combination with its very weak dependence on $\Delta$ in the non-equilibrated regime is highly favorable for using the downstream noise as a thermometer for measuring $T_m$. Indeed, even if the degree of thermal equilibration and the value of $\Delta$ are not known, one can use one of the limiting formulas (for the full equilibration or no equilibration) and obtain in any case a result for $T_m$ with a reasonable accuracy.

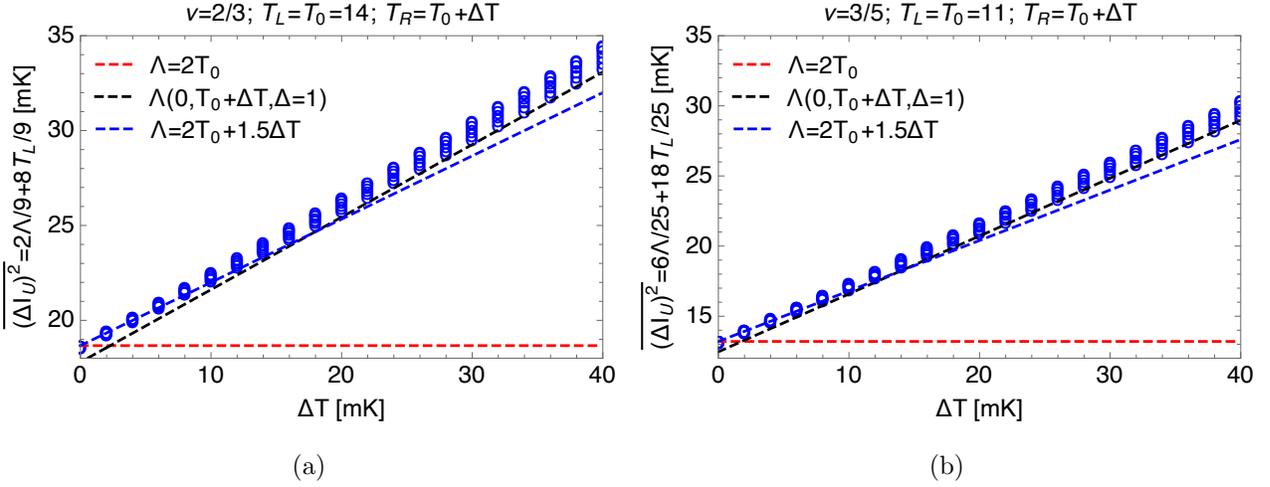

FIG. S9. Upstream noise, $\overline{(\Delta I_U)^2}$, as a function of $\Delta T = T_R - T_0$ [see Eqs. (S49) and (S51)] in units of mK. We set $e^2/h = k_B = 1$ so that $\overline{(\Delta I_U)^2}$ also is in units of mK. (a) For $\nu = 2/3$, with the base temperature $T_0 = 14$, corresponding to the measured $T_0 = 11$ mK at this filling. The blue circles correspond to $\Delta$ between 1 and 2. The blue dashed line, Eq. (S50), is an excellent approximation for $0 \lesssim \Delta T \lesssim 30$. The black dashed line is the large-$\Delta T$ asymptotics $\Lambda(0, \Delta T + T_0, \Delta = 1) \simeq 1.72(T_0 + \Delta T)$. The red dashed line is the Johnson-Nyquist noise with the temperature $T_0$ [see Eq. (S44)], for which $\overline{(\Delta I_m)^2} = 4T_0/3$; this corresponds to the limit of full thermal equilibration. (b) Similar to (a) but at $\nu = 3/5$, with $T_0 = 11$ corresponding to the measured $T_0 = 11$ mK at this filling. The red, dashed line is Johnson-Nyquist noise $\overline{(\Delta I_m)^2} = 2G_{2T}T_0$ with $G_{2T} = 3/5$.

Turning next to $\nu = 3/5$, we have from Eq. (S45) that the downstream fluctuations become

$$\overline{(\Delta I_m)^2} = \frac{6}{25}\frac{e^2}{h} \times \Lambda(\Delta T + T_0, T_0, \Delta) + \frac{18}{25}\frac{e^2}{h}k_B(\Delta T + T_0). \tag{S48}$$

This is plotted in Fig. S8b for $T_0 = 11$, since the base temperature at $\nu = 3/5$ was measured to 11 mK. The conclusions of the analysis is the same as those for $\nu = 2/3$.

## S9. CALCULATION OF UPSTREAM NOISE

In this Section, we use the results of Sec. S7 to compute the upstream noise on an edge segment. The upstream noise configuration corresponds to the choice $T_R = T_m = T_0 + \Delta T$ and $T_L = T_0$ with $\Delta T > 0$. With this choice of parameters, we identify Eq. (S42) with the upstream noise $\overline{(\Delta I_U)^2}$ from Eq. (S15) and obtain for $\nu = 2/3$

$$\overline{(\Delta I_U)^2} = \frac{2}{9}\frac{e^2}{h} \times \Lambda(T_0, \Delta T + T_0, \Delta) + \frac{8}{9}\frac{e^2}{h} k_B T_0. \tag{S49}$$

We plot $\overline{(\Delta I_U)^2}$ as a function of $\Delta T$ for various $\Delta$ in Fig. S9a. We note that similar to the downstream noise, the $\Delta$-dependence is very weak, especially for small $\Delta T$. In the interval $0 \lesssim \Delta T \lesssim 30$, the perturbative expansion in $\Delta T$,

$$\Lambda(T_0, T_0 + \Delta T, \Delta) \approx 2T_0 + 1.5\Delta T, \tag{S50}$$

which is shown by blue dashed line, serves as an excellent approximation for $\Lambda(T_0, T_0 + \Delta T, \Delta)$. For $\Delta T \gtrsim 30$, the large-$\Delta T$ asymptotics, $\Lambda(0, T_0 + \Delta T, \Delta = 1) \simeq 1.72(T_0 + \Delta T)$ (black dashed line) is a better approximation. Note that both asymptotics are actually rather close to each other and in fact serve as very good approximations in the whole range of $\Delta T$.

Comparing Figs. S8 and S9, we see that, for given $\Delta T$, the upstream noise $\overline{(\Delta I_U)^2}$ is weaker than the downstream noise $\overline{(\Delta I_m)^2}$. The main reason is the second term in Eq. (S18) which corresponds to "hot fluctuations" for the downstream noise but to "cold fluctuations" for the upstream noise. This asymmetry between downstream and upstream "$\Delta T$ noise" is a consequence of the chiral nature of the edge.

A similar analysis for $\nu = 3/5$ gives, using Eq. (S18), the upstream noise

$$\overline{(\Delta I_U)^2} = \frac{6}{25}\frac{e^2}{h} \times \Lambda(T_0, \Delta T + T_0, \Delta) + \frac{18}{25}\frac{e^2}{h} k_B T_0. \tag{S51}$$

This is plotted for $\nu = 3/5$ in Fig. S9b. Similarly to $\nu = 2/3$, the approximation (S50) is excellent in the considered temperature regime.

## S10. POWER BALANCE EQUATION AND THEORETICAL FOUNDATIONS OF EXPERIMENTAL DETERMINATION OF HEAT CONDUCTANCE.

In this Section, we present theoretical foundations for experimental determination of the heat conductance $\kappa_{2T}$ on the basis of measurements of the noises $S_{\text{excess}}^{\text{D}}$ and $S_{\text{excess}}^{\text{U}}$ as functions of the heating current $I$.

The heat conductance $\kappa_{2T}$ of the device is obtained by the principle developed in Ref. S31 and applied to FQH states in Refs. S14, S15, and S32. The key principle at work is that, in the steady state, a power $P_{\text{diss}}$ is dissipated in $\Omega_m$, which in turn is evacuated by heat currents carried by phonons, $J^Q_{\text{phonons}}$, and edge states, $J^Q_{\text{edge}}$. This relation is captured by the power balance equation

$$P_{\text{diss}} = J^Q_{\text{edge}} + J^Q_{\text{phonons}} \approx J^Q_{\text{edge}} = \frac{3\kappa_{2T}}{2}(T_m^2 - T_0^2). \quad \text{(S52)}$$

Here, it is assumed that $J^Q_{\text{phonons}} \propto T_m^5 - T_0^5$ can be neglected, which is the case for sufficiently low temperatures. (Alternatively, $J^Q_{\text{phonons}}$ can be independently measured and accounted for.) The edge heat current $J^Q_{\text{edge}}$ includes heat evacuated on all three connected arms. Equation (S52) shows that $\kappa_{2T}$ (in units of $\kappa_0 = \pi^2 k_B^2/3h$) can be extracted from a plot of $P_{\text{diss}}$ vs $(T_m^2 - T_0^2)$: it is given by $2/(3\kappa_0)$ times the slope of the corresponding linear dependence. As explained in detail in the preceding Sections, $T_m$ and $T_0$ are found from noise measurements [see Eqs. (S12), (S46), and (S48)]. The remaining ingredient for determining experimentally the heat conductance is the heat current $J^Q_{\text{edge}}$ equal to the power $P_{\text{diss}}$ dissipated at the central contact $\Omega_m$. We explain now how the value of $P_{\text{diss}}$ is found.

For states with only downstream heat flow (e.g., particle-like or integer edges), $P_{\text{diss}}$ is simply equal to the injected power $P$ [given by Eq. (S6)], $P_{\text{diss}} = P$, since all injected energy is transported downstream to $\Omega_m$ (assuming small bulk losses)[S31]. By contrast, for states with upstream heat flow, such as $\nu = 2/3$ and $\nu = 3/5$, part of the dissipated Joule heat in a hot spot may propagate upstream without heating the central contact. In this case, the power dissipated at $\Omega_m$ may be smaller than the total injected power[S33], $P_{\text{diss}} < P$. In order to use Eq. (S52) for the determination of $\kappa_{2T}$, it is thus important to explore what part of the injected power is actually used to heat the contact $\Omega_m$, i.e., to calculate the ratio $P_{\text{diss}}/P$. In the following, we perform this analysis in the regime of vanishing thermal equilibration.

The injected electrical power on each of the arms $i = 1, 2$ is $P_i = I^2/(2G_{2T})$. This power is dissipated not only inside $\Omega_m$ but also within a finite region of size $l^C_{\text{eq}}$ outside $\Omega_m$, i.e., at the hot spot (see Fig. S10). The power dissipated at this hot spot is[S12]

$$P_{i,\text{h.s.}} = \frac{I^2}{2G_{2T}} \times \frac{\nu_-}{\nu_+}. \quad \text{(S53)}$$

For $\nu_+ = 1$, which holds for all hole-conjugate states in the lowest Landau level (particularly for $\nu = 2/3$ and $\nu = 3/5$ considered in this work), we have $\nu_- = 1 - \nu$ and Eq. (S53) simplifies to

$$P_{i,\text{h.s.}} = \frac{I^2}{2G_{2T}}(1 - \nu). \tag{S54}$$

In the absence of thermal equilibration, a fraction of this power propagates ballistically away from $\Omega_m$ (upstream), while the remaining part propagates to $\Omega_m$ and thus contributes to $P_{\text{diss}}$. In the extreme case where $P_{i,\text{h.s.}}$ is distributed uniformly over the edge modes, we have that only a part of $P_{i,\text{h.s.}}$ propagates downstream and heats $\Omega_m$:

$$P_\Omega = \frac{n_d}{n_d + n_u} \times P_{i,\text{h.s.}}. \tag{S55}$$

Here, $n_d$ and $n_u$ are the numbers of downstream and upstream edge channels, respectively. The remaining contribution, $\frac{n_u}{n_d+n_u} \times P_{i,\text{h.s.}}$ is instead carried by upstream propagating modes away from $\Omega_m$, i.e., back towards the source contact. The electrical power dissipated directly inside $\Omega_m$ is given by the voltage drop close to $\Omega_m$. This directly dissipated power is given by

$$P_{\text{dir}} = \nu \frac{I^2}{2G_{2T}}. \tag{S56}$$

Conservation of energy is ensured by $2P_{i,\text{h.s.}} + 2P_\Omega = P$. Adding the contributions (S55) and (S56), for arms 1 and 2, we obtain for the total dissipated power on $\Omega_m$:

$$P_{\text{diss}} = 2P_{\text{dir}} + 2P_m = \frac{I^2}{G_{2T}}\left[\nu + \frac{n_d}{n_d + n_u}(1 - \nu)\right] = \frac{I^2}{G_{2T}}\left[\frac{n_d + n_u\nu}{n_d + n_u}\right] \equiv \beta \frac{I^2}{G_{2T}}. \tag{S57}$$

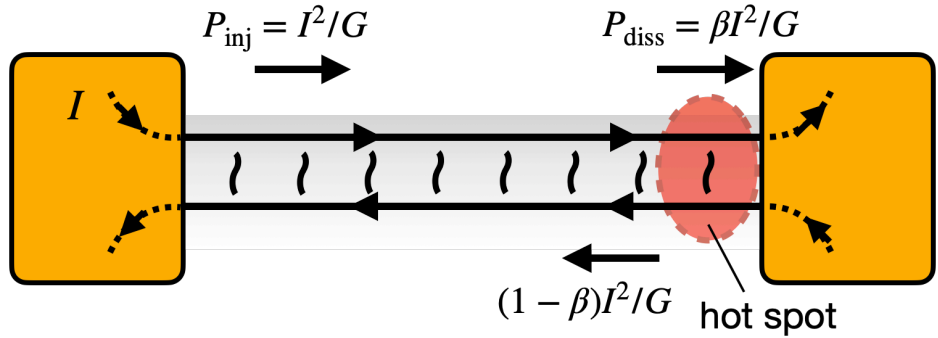

FIG. S10. Dissipation of injected power. Upon injecting a current in the left contact, only a fraction $\beta$ of the injected power $P_{\text{diss}}$ is dissipated in the right contact if the edge channels are not thermally equilibrated. The fraction $(1 - \beta)$ is carried upstream and does not heat the right contact.

Equation (S57) determines the fraction $\beta$ of the injected power that is used to heat $\Omega_m$ in the absence of thermal equilibration. For $\nu = 2/3$ we have $n_u = n_d = 1$, which yields $\beta = 5/6 \approx 0.83$. For $\nu = 3/5$, with $n_d = 1$ and $n_u = 2$, we get $\beta = 11/15 \approx 0.73$. For an edge without counterpropagating modes ($n_u = 0$), we have $\beta = 1$, so that the dissipated power is equal to the injected power, as expected.

Using Eq. (S57), we can rewrite the power balance equation (S52) in the form

$$\frac{I^2}{G_{2T}} = \frac{3\kappa_{2T}}{2\beta}(T_m^2 - T_0^2). \tag{S58}$$

According to Eq. (S58), the heat conductance can be obtained by plotting $P$ vs $(T_m^2 - T_0^2)$. The dimensionless heat conductance $\kappa_{2T}/\kappa_0$ is then given by $2\beta/(3\kappa_0)$ times the slope of the resulting linear dependence.

The calculation of $\beta$ in Eq. (S57) is using a specific model of the contact and hot spot, and hence the actual value can differ somewhat. Also, let us emphasize that $\beta$ is quite close to unity. Previous works[S14,S32] used $\beta = 1$ and obtained proper values for the equilibrated regime. In view of this, and since we do not have experimental control on $\beta$, we also choose $\beta = 1$ in our analysis (i.e., we assume that all dissipated power heats the central ohmic contact).

## S11. HEAT CONDUCTANCE: THEORY

In this Section, we discuss the expected value of the heat conductance $\kappa_{2T}$. The heat conductance for $\nu = 2/3$ was analyzed in various regimes in Ref. S6, and we briefly sketch the relevant results. As in the rest of this work, we focus on the regime of negligible thermal equilibration, $L \ll l_{\text{eq}}$, where $l_{\text{eq}}$ is the thermal equilibration length. As shown in Ref. S6, this regime is subdivided into two. For very short systems (or very low temperatures), $L \ll L_T$, where $L_T$ is the thermal length,

$$L_T = \frac{1}{(v_+^{-1} + v_-^{-1})T}, \tag{S59}$$

the two-terminal heat conductance $\kappa_{2T}$ is given (up to a small correction) by the number of edge modes,

$$\frac{\kappa_{2T}}{\kappa_0} = 2, \qquad L \ll L_T. \tag{S60}$$

For larger $L$ (or higher $T$), the heat conductance is reduced due to partial back-scattering of thermal bosons at boundaries between the non-interacting contacts and interacting part of the wire, cf. Sec. S7:

$$\frac{\kappa_{2T}}{\kappa_0} = 2 - \frac{4R}{1+R} = 2\sqrt{\frac{1}{7\Delta^2 - 4\sqrt{3\Delta^2(\Delta^2-1)} - 3}}, \qquad L_T \ll L \ll l_{\rm eq}. \qquad (S61)$$

In the non-interacting limit, $\Delta = 2$, we have $R = 0$ (no reflection of bosons at the interfaces), and Eq. (S61) yields $\kappa_{2T}/\kappa_0 = 2$. On the other hand, for $\Delta = 1$, corresponding to a regime of strong interaction, one finds

$$\frac{\kappa_{2T}}{\kappa_0} = 1, \qquad L_T \ll L \ll l_{\rm eq}, \qquad (S62)$$

i.e., the thermal conductance us reduced by a factor of two. For the three-arm geometry of the present work, we obtain, for the regime of vanishing thermal equilibration and for strong interactions ($\Delta \approx 1$), a total heat conductance

$$3\frac{\kappa_{2T}}{\kappa_0} = 3. \qquad (S63)$$

The scattering analysis can be generalized to other hole-conjugated FQH edges (at fillings $\nu = p/(2p-1)$ with integer $p > 1$) with counter-propagating edge channels. The analysis is simplified at a low-energy fixed point where neutral modes are completely decoupled with a charge mode; this fixed point corresponds to $\Delta = 1$ for $\nu = 2/3$, discussed above. At such a fixed point, the neutral sector possesses a (global) $SU(p)$ symmetry[S34,S35]. Since this fixed point is a basin in a wide range of interaction-parameters space, it would be desirable to understand the value of thermal conductance at the fixed point.

We consider the case of $\nu = 3/5$ ($p = 3$) for a detailed computation. The edge of the $\nu = 3/5$ hosts one downstream mode $\phi_1$ and two upstream modes $\phi_{1/3}$ and $\phi_{1/15}$, which satisfy the commutation relation $[\phi_1(x), \phi_1(x')] = i\pi\,{\rm sgn}(x-x')$, $[\phi_{1/3}(x), \phi_{1/3}(x')] = -i\pi\,{\rm sgn}(x-x')$, and $[\phi_{1/15}(x), \phi_{1/15}(x')] = -i\pi\,{\rm sgn}(x-x')$, respectively; furthermore $[\phi_j(x), \phi_{j'}(x')] = 0$ for $j \neq j'$ with $j, j' = 1, 1/3, 1/15$. We define a downstream charge mode ($\phi_c$) and two neutral modes ($\phi_{n_1}$ and $\phi_{n_2}$) in terms of $\phi_1$, $\phi_{1/3}$, and $\phi_{1/15}$ as

$$\phi_c = \sqrt{\frac{5}{3}}\left(\phi_1 + \frac{\phi_{1/3}}{\sqrt{3}} + \frac{\phi_{1/15}}{\sqrt{15}}\right),$$

$$\phi_{n_1} = \frac{1}{\sqrt{2}}(\phi_1 + \sqrt{3}\phi_{1/3}),$$

$$\phi_{n_2} = \frac{1}{\sqrt{6}}\left(\phi_1 + \frac{\phi_{1/3}}{\sqrt{3}} + \frac{10}{\sqrt{15}}\phi_{1/15}\right). \qquad (S64)$$

These charge and neutral modes are eigenmodes at the $SU(3)$ symmetric low-energy fixed point, and are thus decoupled.

We next calculate the thermal conductance employing the contact model depicted in Fig. S7. This is the model used for the noise calculation discussed in Sec. S7. The edge segment consists of three regions: the left contact region, a central region, and the right contact region. While the bare modes $\phi_1$, $\phi_{1/3}$, and $\phi_{1/15}$ are incoming modes out of the contacts or outgoing modes to the contacts, the charge and neutral modes are assumed to be eigenmodes in the central region. The left contact is thermally biased compared with the left contact: $T_L > T_R$. Scattering of the bosonic modes between different regions is described by the scattering matrices $S_L$ and $S_R$, given as

$$\begin{pmatrix} \phi_c \\ \phi_{1/3,L} \\ \phi_{1/15,L} \end{pmatrix} = S_L \begin{pmatrix} \phi_{1,L} \\ \phi_{n_1} \\ \phi_{n_2} \end{pmatrix}, \quad \begin{pmatrix} \phi_{1,R} \\ \phi_{n_1} \\ \phi_{n_2} \end{pmatrix} = S_R \begin{pmatrix} \phi_c \\ \phi_{1/3,R} \\ \phi_{1/15,R} \end{pmatrix}, \tag{S65}$$

with

$$S_L = \begin{pmatrix} \sqrt{\frac{3}{5}} & \sqrt{\frac{3}{10}} & \sqrt{\frac{1}{10}} \\ -\sqrt{\frac{1}{3}} & \sqrt{\frac{2}{3}} & 0 \\ -\sqrt{\frac{1}{15}} & -\sqrt{\frac{1}{30}} & \frac{3}{\sqrt{10}} \end{pmatrix}, \quad S_R = S_L^{-1}. \tag{S66}$$

Here, $\phi_{j,L}$ and $\phi_{j,R}$ for $j = 1, 1/3, 1/15$ are incoming or outgoing modes in the left ($L$) and the right $R$ contacts, respectively. This type of scattering problem is characterized by transmission amplitude $\mathcal{T}(\omega)$ from the left to the right, calculated from the scattering matrices $S_L$ and $S_R$ as

$$\mathcal{T}(\omega) = \frac{T}{1 - Re^{2i\omega\tilde{\tau}}} = \frac{\nu}{1 - (1-\nu)e^{2i\omega\tilde{\tau}}} \tag{S67}$$

with $T = \nu$ and $R = (1 - \nu)$ being the the transmission and reflection coefficients for the $\phi_1$ mode. Furthermore, $\tilde{\tau} = L(1/v^+ + 1/v^-)/2$ is the mean flight time through the central region, $v_+$ and $v_-$ are the velocities of the charge mode and the neutral modes, respectively. Note that the $SU(3)$ symmetry of the neutral sector renders the velocities of the two neutral modes to be the same. Calculating the energy current in the edge mode $\phi_1$, we obtain the reflected thermal conductance $\kappa_{12}$ with

$$\kappa_{12} = \kappa_0 \left(1 - \frac{6}{\pi^2 \bar{T}^2} \int_0^\infty \frac{\omega d\omega}{e^{\omega/\bar{T}} - 1} |\mathcal{T}(\omega)|^2\right). \tag{S68}$$

Here, $\bar{T}$ is the average temperature $\bar{T} = (T_L + T_R)/2$. For very short length $L \ll L_T \equiv 1/[(v_+^{-1} + v_-^{-1})\bar{T}]$, $\kappa_{12}$ becomes zero, and hence the two-terminal heat conductance $\kappa_{2T} = (n_u + n_d)\kappa_0 - 2\kappa_{12}$ is given by the total number of modes $\kappa_{2T} = (n_u + n_d)\kappa_0$. On the other hand, in the regime of $L_T \ll L \ll l_{\text{eq}}$, we have $\kappa_{12} = \kappa_0 \left(2R/(1+R)\right)$ and thus

$$\kappa_{2T} = \kappa_0 \left( n_u + n_d - \frac{4R}{1+R} \right) = \kappa_0 \left( n_u + n_d - \frac{4(1-\nu)}{2-\nu} \right). \tag{S69}$$

Note that Eqs. (S67)-(S69) generically holds for any hole-conjugated states with $\nu = p/(2p-1)$ at a $SU(p)$ fixed point. The case $p = 2$ corresponds to $\nu = 2/3$, with $n_u = n_d = 1$. Equation (S69) yields in this case $\kappa_{2T} = \kappa_0$, in agreement with Eq. (S62). For $p = 3$ we have $\nu = 3/5$, with $n_u = 2$ and $n_d = 1$, and Eq. (S69) yields

$$\frac{\kappa_{2T}}{\kappa_0} = 13/7 \approx 1.86, \tag{S70}$$

as stated in the main text. The total heat conductance of the three-armed devices is then obtained as $3\kappa_{2T}/\kappa_0 \approx 5.57$.

## S12. EXTRACTING HEAT CONDUCTANCE AND TEMPERATURE DEPENDENCE OF UPSTREAM NOISE FROM EXPERIMENTAL DATA AND COMPARISON TO THEORY

In this Section, we present details of determining the heat conductance and the temperature dependence of the upstream noise from the experimental data by using the theoretical framework developed in Secs. S6-S11. Further, we compare the obtained results for the upstream noise and the heat conductance to the theoretical predictions for the FQH state with filling factor $\nu = 2/3$.

### A. Determining the central-contact temperature $T_m$

As a first step, we extract the central contact temperature $T_m$ for a given injected current $I$ (or equivalently a given injected power $P$). For this purpose, we combine Eq. (S12) for the downstream excess noise $S_{\text{excess}}^{\text{D}}$ with the microscopically computed $\overline{(\Delta I_m)^2}$ from Eqs. (S46) and (S48). We further use the approximation (S47). As a result, we obtain the following equation relating $\Delta T = T_m - T_0$ and the measured noises:

$$\frac{4\alpha}{3} \times G_{2T} k_B \Delta T \approx S_{\text{excess}}^{D} - \frac{1}{9} \left[ S_{\text{excess}}^{S_1} + S_{\text{excess}}^{S_2} + S_{\text{excess}}^{U} \right], \quad G_{2T} = \frac{\nu e^2}{h}. \tag{S71}$$

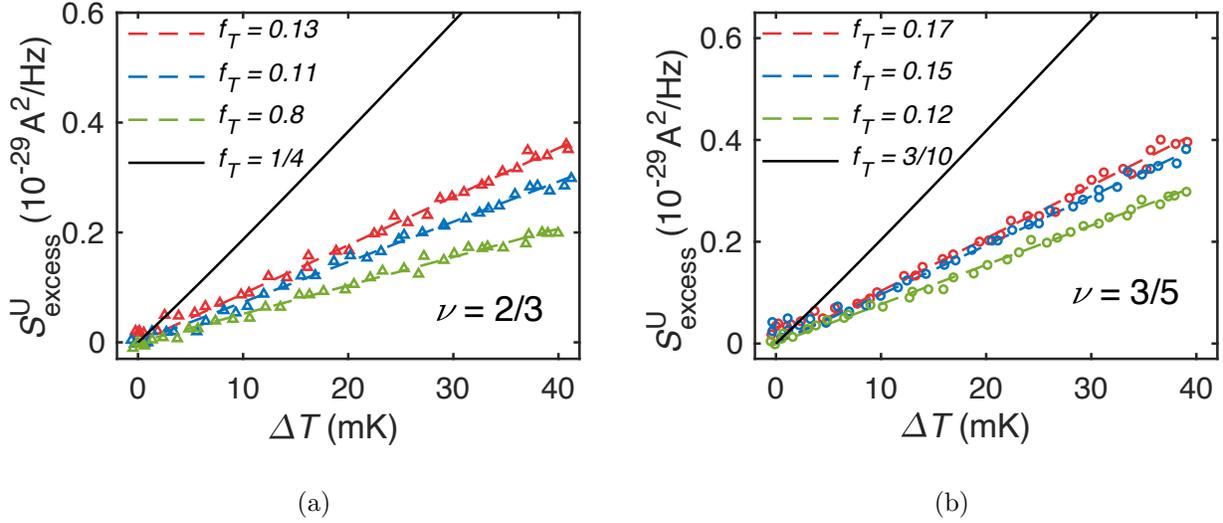

FIG. S11. Upstream excess noise $S^{\rm U}_{\rm excess}$ vs $\Delta T = T_m - T_0$, where $T_m$ is the temperature of the hot central contact and $T_0$ is the base temperature. (a) Filling $\nu = 2/3$. Triangle symbols: experimentally measured noise, with the temperature $\Delta T$ determined as explained in Sec. S12 B, for three different lengths for the upstream heat propagation: $15\mu$m (red), $45\mu$m (blue), and $75\mu$m (green). Dashed lines are linear fits, which gives thermal Fano factors $f_T \equiv S^{\rm U}_{\rm excess}/(2G_{2T}k_B\Delta T)$. The black solid line is the theoretical result (for vanishing thermal equilibration and under assumption of no losses to the bulk), $f_T = 1/4$ [see Eq. (S72)], which becomes $S^{\rm U}_{\rm excess} \approx 0.017\Delta T$ when the noise is measured in units of $10^{-29}$A$^2$/Hz and the temperature in mK. (b) Filling $\nu = 3/5$. Circles denote measured data with color coding as that in (a). The black line is the theoretical result (S73): $f_T = 3/10$ or $S^{\rm U}_{\rm excess} \approx 0.019\Delta T$ in the units referred to in (a).

Here, $\alpha = (4\nu_+ - 3\nu_-)/\nu_+$, which yields $\alpha = 3/4$ for $\nu = 2/3$ and $\alpha = 7/10$ for $\nu = 3/5$. Furthermore, $S^{\rm D}_{\rm excess}$ and $S^{\rm U}_{\rm excess}$ are measured directly for various lengths between $\Omega_m$ and the upstream contact U with the amplifier $A_{\rm U}$ (see Fig. S6). By contrast, the noise from sources, $S^{S_1}_{\rm excess}$ and $S^{S_2}_{\rm excess}$, are not directly measured. However, since these noises are of exactly the same nature as the upstream noise $S^{\rm U}_{\rm excess}$, they should be essentially equal to $S^{\rm U}_{\rm excess}$ at lengths equal to the distances between the sources and $\Omega_m$. In the experiment, these lengths were fixed at $30\mu$m and $150\mu$m for $S_1$ and $S_2$, respectively. Thus, all the terms in the r.h.s. of Eq. (S71) are obtained from experimental measurements, which allows us to determine $\Delta T$ as a function of $I$.

## B. Determination of temperature dependence of upstream noise. Comparison to theory.

Next, we plot the measured $S_{\text{excess}}^{\text{U}}$ against $\Delta T$ extracted from Eq. (S71) as explained in Sec. S12 A. The results are shown in Fig. S11 for three different values of the length between $\Omega_m$ and the upstream contact U. While the noise decreases with increasing length, we see that this dependence is rather weak. This is particularly true for shorter systems: the noise decreases only by $\approx 10\%$ when the length increases from $15\mu$m to $45\mu$m. This is a clear demonstration of the fact that the system is in the regime with essentially vanishing heat equilibration: the heat propagates ballistically from the central contact $\Omega_m$ to the noise spot near the contact U. A slow reduction of noise with increasing length may be due to two reasons: (i) weak thermal equilibration, and (ii) losses of heat propagating along the edge to the environment ("bulk"), including phonons as well as electronic modes in the bulk mediated by Coulomb interaction[S36]. As the data on heat conductance demonstrate (see Sec. S12 C below), the dominant source in our experiment is losses to the bulk, while the effect of thermal equilibration within the edge is negligible.

In Fig. S11 we further compare the results for $S_{\text{excess}}^{\text{U}}(\Delta T)$ (obtained from experimental measurements as detailed above) to the theoretical formula for the upstream noise in the regime of vanishing thermal equilibration,

$$S_{\text{excess}}^{\text{U}} = \frac{1}{3}\frac{e^2}{h} k_B \Delta T \equiv \frac{1}{4} \times 2 G_{2T} k_B \Delta T, \qquad G_{2T} = \frac{2e^2}{3h}. \tag{S72}$$

$$S_{\text{excess}}^{\text{U}} = \frac{9}{25}\frac{e^2}{h} k_B \Delta T \equiv \frac{3}{10} \times 2 G_{2T} k_B \Delta T, \qquad G_{2T} = \frac{3e^2}{5h}. \tag{S73}$$

Equation (S72) is obtained from the general expression (S15) of $S_{\text{excess}}^{\text{U}}$ by inserting Eq. (S49) and using the approximation (S50). Similarly, Eq. (S73) follows by using Eqs. (S50) and (S51) in Eq. (S15). Equations (S72) and (S73) yield the excess upstream noise in the limit of zero thermal equilibration and no losses to the bulk, i.e., when the heat from $\Omega_m$ propagates ballistically and without losses to the noise spot near the contact U (see Fig. S6).

Equations (S72)-(S73) define the thermal Fano factors $f_T \equiv S_{\text{excess}}^{\text{U}}/(2 G_{2T} k_B \Delta T)$, which for $\nu = 2/3$ and $\nu = 3/5$ gives $f_T = 1/4$ and $f_T = 3/10$ respectively.

We see that the linear dependencies of $S_{\text{excess}}^{\text{U}}(\Delta T)$ is in full consistency with experimental data. At the same time, the experimental value of the noise for the shortest distance (for which the effect of losses and thermal relaxations are negligible as pointed out above) is approximately twice smaller than the theoretical prediction. It is not clear to us at present what is the source of this discrepancy.

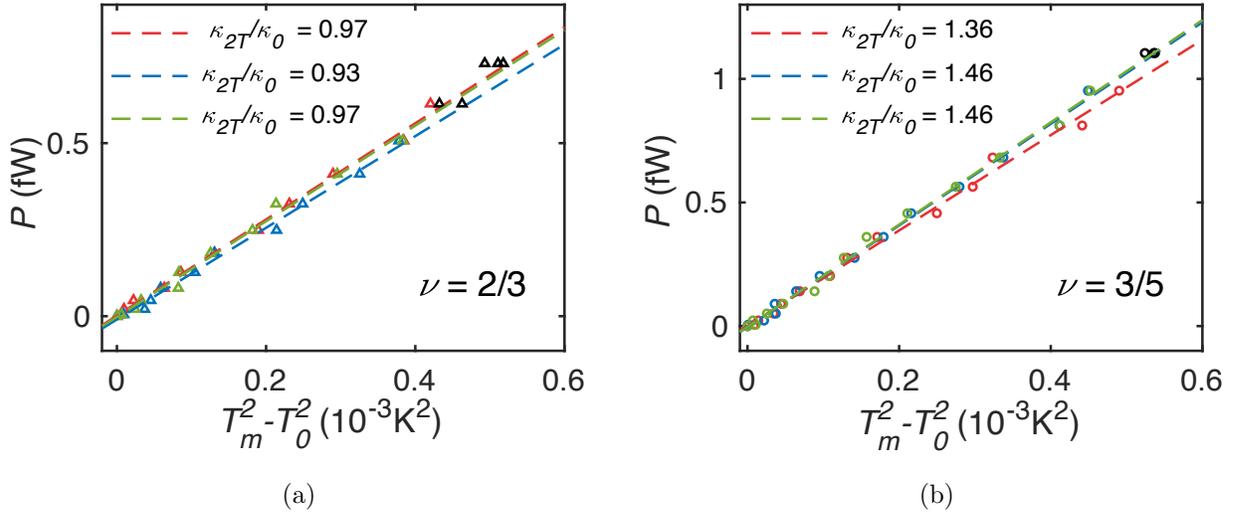

FIG. S12. Injected power $P$ vs $T_m^2 - T_0^2$. (a) Filling $\nu = 2/3$. Triangles denote measured data for three lengths between the central contact $\Omega_m$ and the upstream contact U: 15$\mu$m (red), 45$\mu$m (blue), and 75$\mu$m (green). The heat conductance $\kappa_{2T}/\kappa_0$ is given by the slope of the dependence $P$ vs $T_m^2 - T_0^2$, multiplied by $2/(3\kappa_0)$, [see Eq. (S57), were we take $\beta = 1$.]. This results in $\kappa_{2T}/\kappa_0 \approx 0.97$, 0.93, and 0.97 for the lengths 15$\mu$m, 45$\mu$m, and 75$\mu$m, respectively. (b) Filling $\nu = 3/5$. Circles denote measured data at the same lengths as in (a). We find $\kappa_{2T}/\kappa_0 \approx 1.36$, 1.46, and 1.46 for the lengths 15$\mu$m, 45$\mu$m, and 75$\mu$m, respectively.

### C. Determination of heat conductance. Comparison to theory.

To extract the heat conductance $\kappa_{2T}/\kappa_0$ from the experimental measurements, we use the approach described in Sec. S10. In accordance with the power balance equation (S58), we plot in Fig. S12 the injected power $P$ as a function of $T_m^2 - T_0^2 = \Delta T^2 + 2T_0 \Delta T$. Here, the base temperatures $T_0 = 14$mK for $\nu = 2/3$ and $T_0 = 11$mK for $\nu = 3/5$ are extracted from equilibrium noise (i.e., noise at zero current bias $I = 0$). The temperature difference $\Delta T$ is obtained from the experimentally measured noise according to Eq. (S71). We extract the heat conductance for three lengths, 15$\mu$m (plotted in red), 45$\mu$m (blue), and 85$\mu$m (green), between $\Omega_m$ and the upstream contact U. The heat conductance $\kappa_{2T}/\kappa_0$ is obtained as $2/\kappa_0$ times the slope of the resulting linear dependences, see Eq. (S58) (as motivated in Sec. S10, we have taken $\beta = 1$ for all considered states). The slopes are extracted in the linear regime $0 \lesssim T_m^2 - T_0^2 \lesssim 0.4 \times 10^{-3} \mathrm{K}^2$. For $\nu = 2/3$, we find $\kappa_{2T}/\kappa_0 \approx 0.97$, 0.93, and 0.97 for the lengths 15$\mu$m, 45$\mu$m, and 75$\mu$m, respectively. For $\nu = 3/5$, we find $\kappa_{2T}/\kappa_0 \approx 1.36$, 1.46, and 1.46 for the lengths 15$\mu$m, 45$\mu$m, and 75$\mu$m, respectively. The result is that $\kappa_{2T}$ independent on the length, within the uncertainty determined by the statistical scattering of data.

The length-independence of the measured heat conductance is a strong evidence of the absence (within the accuracy of our data) of heat equilibration at the length scales studied in our experiment. Indeed, in the presence of heat equilibration, the heat conductance $\kappa_{2T}$ of a $\nu = 2/3$ edge shows, with increasing length $L$, a crossover from the ballistic behavior ($L$-independent $\kappa_{2T}$) at $L \ll l_{\mathrm{eq}}$ to the diffusive behavior, $\kappa_{2T} \propto 1/L$ at $L \gg l_{\mathrm{eq}}$, see Ref. S6. Our results thus imply that the devices studied in this experiment are in the regime $L \ll l_{\mathrm{eq}}$.

Let us further comment on the insensitivity of the thermal-conductance measurements to losses to the bulk. Indeed, for the length 75$\mu$m, these losses are clearly observable in the upstream noise measurement, see Fig. S11. At the same time, the results for the thermal conductance shown in Fig. S12 do not show any trace of the losses. The reason for this insensitivity of the heat conductance measured with the present method to bulk losses is as follows[S33]. The measurement protocol matches the incoming and outgoing heat flows with respect to $\Omega_m$ [see Eq. (S58)]. The crucial point is that the heat is evacuated from the central contact $\Omega_m$ via the edge states. Whether the full outgoing energy reaches another electrode or some part of it leaks to the bulk on the way there is irrelevant. The assumption here is that the heat that is leaking to the bulk does not return to the central contact $\Omega_m$, which is expected to be a very good approximation. This should be contrasted to the effect of heat equilibration within the edge, which leads to back-scattering of heat that thus returns to $\Omega_m$, leading to a decrease of $\kappa_{2T}$.

Summarizing, our results for the upstream noise and the conductance are interpreted in the following way: (i) the thermal relaxation within the edge is negligible at studied length scales, and (ii) leakage of energy to the bulk leads to reduction of the upstream noise with distance but is irrelevant for the thermal conductance.

Finally, we compare the value $\kappa_{2T}/\kappa_0 \approx 1$ obtained from experimental data at $\nu = 2/3$ with the theoretical result (S62) derived for the thermally non-equilibrated regime and strong interaction ($\Delta \approx 1$). We see that the experimental value is in good agreement with the theoretical prediction. For $\nu = 3/5$, the obtained $\kappa_{2T}/\kappa_0 \approx 1.45$ lies below the predicted $\kappa_{2T}/\kappa_0 = 13/7 \approx 1.86$ [see Eq. (S70)]. This deviation could be related to a deviation of the system from the infrared fixed point.